\documentclass[letterpaper]{article}

\usepackage[english]{babel}

\usepackage[T1]{fontenc}

\usepackage[letterpaper,top=1in,bottom=1in,left=1in,right=1in,margin=1in]{geometry}

\usepackage{etoolbox}
\usepackage{setspace}
\usepackage{subcaption}
\usepackage{float}
\usepackage{amsmath}
\usepackage{svrsymbols}
\usepackage{titletoc}
\usepackage[titles]{tocloft}
\usepackage{tikz-feynman}
\tikzfeynmanset{compat=1.1.0}
\usepackage{diagbox}

\usepackage{multirow}

\AtBeginEnvironment{quote}{\singlespacing\small}

\usepackage{titlesec}
\titleformat{\section}[block]{\color{black}\Large\bfseries\filcenter}{}{1em}{}

\newcommand*{\BeginNoToc}{%
 \addtocontents{toc}{%
   \edef\protect\SavedTocDepth{\protect\the\protect\value{tocdepth}}%
  }%
  \addtocontents{toc}{%
    \protect\setcounter{tocdepth}{-10}%
  }%
}
\newcommand*{\EndNoToc}{%
  \addtocontents{toc}{%
    \protect\setcounter{tocdepth}{\protect\SavedTocDepth}%
  }%
}

\addto\captionsenglish{
	}
\addto\captionsenglish{
	}
\addto\captionsenglish{
	}
\usepackage{afterpage}

\DeclareRobustCommand*{\contfigures}{%
  \afterpage{{
      {\normalfont\Large\bfseries\centering
 \normalfont LIST OF FIGURES (continued)\par\bigskip}}\vspace{1.5\baselineskip}
      {\normalfont\bfseries{\hspace{-20mm} \normalfont Figure}\hfill \normalfont Page}\vspace{1.2\baselineskip}
  }
}

\begin{document}

\centering
\large{IMPROVEMENT OF THE NOVA NEAR DETECTOR EVENT RECONSTRUCTION AND PRIMARY VERTEXING THROUGH THE APPLICATION OF MACHINE LEARNING METHODS}\\
[8\baselineskip]
\doublespace{
A Thesis by\\
Zakaria A Elkarghli\\
Bachelor of Science, Oklahoma State University 2013}\\
[6\baselineskip]
\singlespacing{
 Submitted to the Department of Mathematics, Statistics, and Physics \\
 and the faculty of the Graduate School of \\
 Wichita State University\\
 in partial fulfillment of\\
 the requirements for the degree of\\
 Master of Science} \\
[10\baselineskip]
December 2020
\pagenumbering{roman}
\thispagestyle{empty}
\pagebreak

\ 
\vspace{3in}
\begin{center}
\textcopyright\doublespacing{
Copyright 2020 by Zakaria Elkarghli \\
All Rights Reserved}
\end{center}
\thispagestyle{empty}
\pagebreak

\begin{center}
IMPROVEMENT OF THE NOVA NEAR DETECTOR EVENT RECONSTRUCTION AND PRIMARY VERTEXING THROUGH THE APPLICATION OF MACHINE LEARNING METHODS\\
[3\baselineskip]
The following faculty members have examined the final copy of this thesis for form and content, and recommend that it be accepted in partial fulfillment of the requirements for the degree of Master of Science, with a major in Physics. \\
[4\baselineskip]
\raggedright{

\rule{3.75in}{0.4pt}\\
Mathew Muether, Committee Chair} \\
[3\baselineskip]
\rule{3.75in}{0.4pt}\\

Ajita Rattani, Committee Member \\
[3\baselineskip]

\rule{3.75in}{0.4pt}\\
Terrance Figy, Committee Member \\
[3\baselineskip]

\rule{3.75in}{0.4pt}\\
Nickolas Solomey, Committee Member \\
[3\baselineskip]

\end{center}
\pagebreak

\begin{center}
DEDICATION \\
\vspace*{2in}
I dedicate this thesis to my father and mother, my brothers and sisters, my friends, and to those aspiring with a purpose greater than themselves.
\end{center}
\pagebreak

\begin{center}
ACKNOWLEDGEMENTS\\
\end{center}
\vspace*{1\baselineskip}
\raggedright{
\setlength{\parindent}{0.50in}
\doublespacing{

I would like to acknowledge and thank this committee for their time and feedback. I would like to particularly thank Prof. Mathew Muether for his mentorship throughout my research period. I am grateful for the Wichita State University physics graduate faculty and staff, not only for the great learning experience, but for sparking a new area of curiosity in the world of science. I would like to sincerely thank the NOvA collaboration group; a special thank you to Dr. Micah Groh who answered many of my technical questions. I would like to thank my family for their encouragement and allowing me to be inquisitive at a young age. I would like to thank those who stood by me in a personal capacity throughout this program - for without them this would not have been possible.

\setlength{\parindent}{10ex}
Finally, I would like to thank Wichita State University for making available to students the high-performance computing cluster, BeoShock, that this work heavily relied upon. This work was supported by the US National Science Foundation grant program.

}
}
\pagebreak

\begin{center}
ABSTRACT\\
\end{center}
\vspace*{1\baselineskip}
\raggedright{
\setlength{\parindent}{0.50in}
\doublespacing{

The purpose of this work is to examine the application of a deep learning model in event reconstruction of neutrino interactions. The challenges faced in event reconstruction include the placement of an accurate primary neutrino interaction vertex which is used to support the particle track and prong algorithms. The result of accurate primary vertex ensures all particles involved in a neutrino interaction are included. We propose a regression-based Convolutional Neural Network (CNN) method to predict the primary vertex of a particle interaction. We show that with raw two-dimensional pixel map views as input, the regression-based CNN can predict the primary vertex in all three coordinates. This work is applied as part of the NOvA (NuMI Off-axis $\nu_e$ Appearance) near detector reconstruction efforts. The primary vertex predicted by the regression-based CNN model shows promising results for future applications. This deep learning method can be extended to secondary vertexing through a Kernel Density Estimate algorithm discussed in this work.

}
}
\pagebreak
\newcommand\Decide[1]{#1}

\tableofcontents
\addtocontents{toc}{{\bfseries \normalfont Chapter\hfill \normalfont Page\bigskip\par}}
\pagebreak

\BeginNoToc
\listoftables
\addtocontents{lot}{{\bfseries \normalfont Table\hfill \normalfont Page\bigskip\par}}
\pagebreak

\listoffigures
\addtocontents{lof}{{\bfseries \normalfont Figure\hfill \normalfont Page\bigskip\par}}
\addtocontents{lof}{\contfigures} 
\pagebreak 
\EndNoToc

\begin{flushleft}

\makeatletter
\def\sectionsuffic{}
\def\subsectionsuffix{\quad}
\def\subsubsectionsuffix{\quad}
\def\paragraphsuffix{\quad}
\renewcommand\@seccntformat[1]{\csname the#1\endcsname\csname#1suffix\endcsname}
\renewcommand\thesection{\protect\Decide{\@arabic\c@section}}
\renewcommand\thesubsection{\@arabic\c@section.\@arabic\c@subsection}
\renewcommand\Decide[1]{}
\makeatother

\allowdisplaybreaks
\begin{center}
\Large\bf \normalfont LIST OF SYMBOLS\\
\end{center}
\begin{center}
\doublespacing{
\begin{align*}
&\nu &&\text{Neutrino}\\
&\overline{\nu}  &&\text{Anti-neutrino}\\
&\nu_e &&\text{Electron Neutrino}\\
&\nu_{\tau} &&\text{Tau Neutrino}\\
&\nu_{\mu} &&\text{Muon Neutrino}\\
&\pi  &&\text{Pion}\\
&\mu  &&\text{Muon}\\
&\tau  &&\text{Tau Lepton}\\
&\gamma  &&\text{Photon/Gamma}\\
&\text{e}  &&\text{Electron}\\
&\text{c}  &&\text{Speed of Light}\\
&\text{E}_{\nu}  &&\text{Neutrino Energy}\\
\end{align*}
}
\end{center}

\pagebreak
\section*{CHAPTER I}
\vspace{0.25in}
\section{Introduction}
\pagenumbering{arabic}
\doublespacing{

\setlength{\parindent}{10ex}
High Energy Physics (HEP) studies the building blocks of matter, antimatter, and their interacting forces. Many of these fundamental particles do not occur naturally but may be created through high energy collisions of parent particles giving this field of physics its name.  By studying HEP, researchers attempt at exploring a wide variety of ideas such as the process of formation of the universe or the Higgs field which gives mass to fundamental particles. More specifically, HEP deals with the study of matter's most elementary particles and their interactions. Elementary particles are defined as particles not found to be compromised of further simpler particles. 

Through research in HEP, neutrons and protons, once thought to be fundamental particles were later found to be comprised of quarks each with varying mass, charge, and spin. Over the following decades, more particles and interactions were later discovered such as positrons, neutrinos, muons and more. With new particles and respective properties discovered, a new framework had to be developed to account for these building blocks of matter. This came to be known as the Standard Model of particle physics\cite{griffits_elem_part}.

The Standard Model is a theoretical structure that not only describes the fundamental particles of matter but also the exchanged particles related to the fundamental forces. It identifies fermions which are the building blocks of matter and bosons which mediate three of the four fundamental forces of nature. One of most familiar force carriers, the photon ($\gamma$) is the force carrier of electromagnetism. The other three force carrier particles, the gluon (g), and the W and Z Bosons, mediate the strong and weak force, respectively. Fermions can further be classified into two groups: leptons and quarks, each with six members of their own. Leptons behave differently than quarks. Most of visible matter is made of protons and neutrons, which are made of quarks. Quarks exist with other quarks as composites, whilst leptons may exist exclusive from  other particles. Three of the six lepton family members have an electrical charge, these are the electron (e), the muon ($\mu$), and the tau ($\tau$). The other three leptons, known as neutrinos ($\nu$), have no charge, a mass close to zero, and have very little interaction with matter\cite{concepts_particle_phys}. 

Neutrinos play a significant role in our understanding of the universe and stellar formation\cite{acero_2020}. A product of nuclear fusion, neutrinos help complete the picture of a star's supernova process. Studying neutrinos is important because a massive neutrino is outside of the Standard Model, which points to a new area in physics\cite{nue_oscillation}. An extensive study of neutrinos and how they interact with other particles must first occur to help broaden humankind's astronomical understanding. Neutrinos are one of the most abundant particles found in the known universe, however, are extremely difficult to detect. To detect and study elementary particles and their respective interactions, HEP researchers have employed the use of particle accelerators.

\begin{figure}[H]
  \centering
    \includegraphics[width=3in]{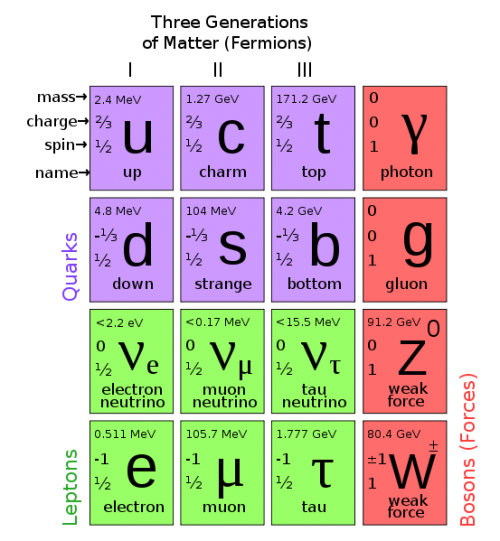}
    \caption[The Standard Model of Particle Physics]{\label{fig:stdmdl} The Standard Model of Particle Physics}
  \end{figure}
\addtocontents{lof}{\vspace{\normalbaselineskip}}

\setlength{\parindent}{10ex}
Particle accelerators typically begin with a proton or an electron and are accelerated using successive electromagnetic fields. As a result of the higher particle momenta, the wavelength is decreased to a degree that they can be directed to collide with a target or an opposing particle beam. High sensitivity detectors are placed strategically to record the following events throughout the collision process. In other instances, absorbers may be used to filter unwanted particles allowing researchers to study a specific daughter particle and its interaction. An instance of this is to study antiparticles. High energy particles impact a specified target generating a new particle and its respective antiparticle. Successive electromagnets and absorbers are then used to separate the antiparticle for further study. 

One such particle research facility is found as a part of the Fermi National Accelerator Laboratory, known as the NOvA (NuMI Off-axis $\nu_e$ Appearance) Project. By colliding protons into graphite targets, researchers produce NuMI (Neutrinos at the Main Injector) muon neutrino beam which is used to study a variety of collisions and associated fundamental particles. Two detectors are utilized in the NOvA experiment: a near detector ($ND$) and a far detector ($FD$), each intended to study different properties of neutrinos. At both detectors, each neutrino interaction is recorded as an event. Detailed event reconstruction is critical for the studies conducted by the NOvA experiment group. 

The NOvA event reconstruction algorithm consists of multiple steps to identify preliminary interaction properties such as global 3D vertex reconstruction, prong formation, event classification, and more. Physical events can be analyzed based on neutrino energy ($E_{\nu}$) derived from accurate event reconstruction. The primary neutrinos of interest to the NOvA group are the muon neutrino ($\nu_{\mu}$) and anti-neutrino ($\overline{\nu}_\mu$), as well as the electron neutrino ($\nu_e$) and anti-neutrino ($\overline{\nu}_e$)\cite{acero_2019}. The study of these neutrinos and their four channels of oscillations contribute to the study of direct CP violation and neutrino mass ordering. The NOvA reconstruction group uses a system based on image recognition techniques known as the Convolutional Visual Network (CVN) to recognize and distinguish neutrino interactions in question. 

The NOvA CVN is based on a trained Convolutional Neural Network (CNN) of multilayer perceptrons (MLP) that can identify and separate neutrino interactions from background noise. The NOvA CVN is based on a custom built GoogLeNet architecture framework named Caffe, developed for deep learning applications\cite{cvnid}. The CVN developed classifications and event pixel maps have allowed for a machine learning based approach to solving problems within the NOvA experiment. The NOvA reconstruction groups for both the far and near detectors have applied deep learning to assist in event slicing, energy reconstruction, prong formation, and particle identification\cite{lars_repo}. The NOvA CVN algorithm has made accessible an array of new machine learning applications through its high-fidelity characteristic finding and event classification.

\begin{figure}[H]
  \centering
    \includegraphics[width=6in]{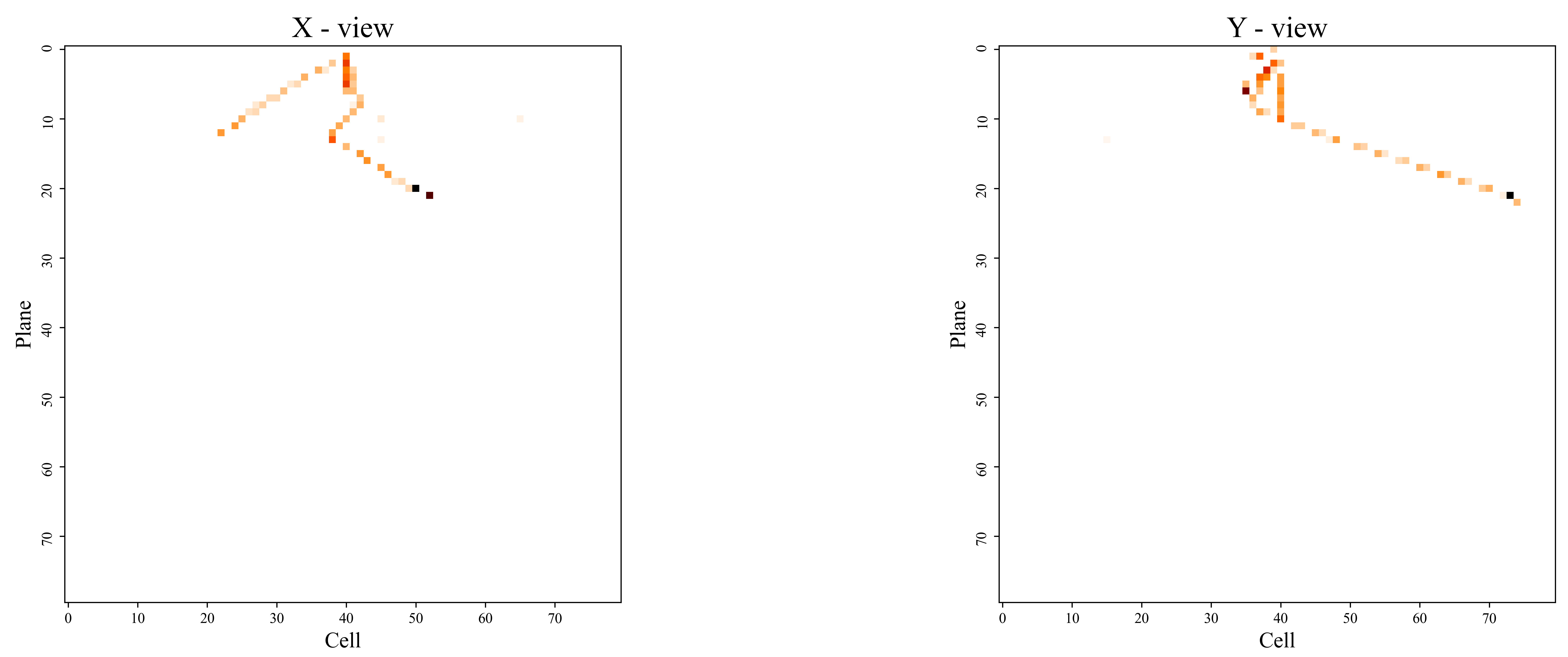}
    \caption[Sample muon neutrino event pixel map pair]{\label{fig:cvnmap} Sample muon neutrino event pixel map pair}
  \end{figure}
\addtocontents{lof}{\vspace{\normalbaselineskip}}

\setlength{\parindent}{10ex}
One area of significance to NOvA researchers is effective hit tracking and vertex detection within an event. An event, within the NOvA detector, is defined as a record of hits within a window of time (500 $\mu$s). Events need further processing to extract or slice independent interactions of interest within one event from background noise. These slices are then clustered in a manner that associate each slice to a neutrino interaction or cosmic ray source. The NOvA experiment records millions of events each year for both the near and far detectors. The need to reliably automate these preliminary steps for a gamut of energies allows research teams more time to focus on the study of interactions in their most complete and purest form. 

The NOvA reconstruction group has used an array of internally developed programs aimed at meeting this objective. Event reconstruction evolved to include the ability for neutrino researchers to plot the hit tracks and the neutrino origin, known as the primary vertex, for each slice. Later, hit tracks were isolated into particle groups, 'prongs' to identify the offshoot of secondary particles generated by the original neutrino collision. The list of reconstruction resources available can be expanded to include models that can predict neutrino energy and identify prong particles, but this will require accurate geospatial metadata\cite{coordconvcvn}. The NOvA reconstruction group has relied on traditional image processing algorithms to track particles in a slice and estimate primary vertices. 

This approach has served the reconstruction group well but like any other algorithm has limitations and biases. More recently, NOvA has sought to address secondary vertexing as another resource for neutrino research groups. As a neutrino collision creates a diverse set of secondary particles, each of these particles generate a set of secondary tracks along their flight path. In some instances, the secondary particles may collide with targets in the near or far detector creating a secondary collision. A reconstruction of secondary events could generate a better understanding of the initial neutrino interaction that occurred within the detector.

The aim of this work is to explore the use of machine learning in the application of vertexing within the NOvA near detector. By utilizing the CVN pixel maps and associated classification data, a supervised regression approach convolutional neural network model was trained to predict primary neutrino vertexing. The learning model utilizes the three-dimensional event topology to predict the location of the primary vertex with a higher degree of accuracy than traditional image processing methods\cite{linear_reg_eucl}. This work also serves to assist in the application of deep learning in secondary vertexing by introducing auxiliary algorithms to address the shortcomings of the learning models. When applied in conjunction with a kernel density estimate-based algorithm, this learning model may be applied to estimate secondary vertex locations for a given slice.

\pagebreak
\section*{CHAPTER II}
\vspace{0.10in}
\section{The NOvA Experiment}
\addtocontents{toc}{\vspace{\normalbaselineskip}}

\setlength{\parindent}{10ex}
The Fermi National Accelerator Laboratory (Fermilab) is a Department of Energy national laboratory focused on the research of high energy particle physics. The Fermilab  complex is a 6,800-acre research facility composed of multiple particle accelerators, detectors, and sensors. Fermilab researchers have developed leading particle accelerators that generate beams of particles used by research groups around the world  aimed at better understanding matter, energy, and the universe. There are over a dozen experiments that are part of the Fermilab’s mission, each focused in studying a specific aspect of particle physics. The focus of this work is Fermilab’s NOvA (NuMI Off-Axis  $\nu_e$ Appearance) experiment which detects and studies neutrinos, antineutrinos, neutrino oscillations, neutrino mass hierarchy, CP violation, and much more.

Neutrino detectors must be designed with a set of requirements aimed at fulfilling certain research criteria. This may include low neutrino energy threshold and resolution to allow for accurate detection and particle identification for neutrino related oscillation studies. Another detector property could include appropriate position, angle and time resolution which would allow researchers to reconstruct events correctly, including particle evolution. No single detector can do everything, which explains the variety of particle physics experiments found at the Fermilab complex. The NOvA experiment was designed as the next evolutionary step based on another neutrino experiment, MINOS (Main Injector Neutrino Oscillation Search)\cite{acero_2019}. 

The main challenge faced with neutrinos is detection; their presence can only be detected if a neutrino interacts with another particle. Neutrino interactions can be classified in two categories: charged-current (CC) and neutral-current (NC) interactions. In charged-current interactions, neutrinos are converted into the respective charged lepton. Charged-current interactions occur through exchanges of $W^{\pm}$ boson\cite{larkoski_elem_part}. Due to the nature and detector signatures of the particles the neutrino is converted to, charged-current interactions are simpler to detect and analyze. As well, charged-current interactions allow researchers the ability to determine the original flavor of the neutrino in a process known as ‘flavor-tagging’\cite{nu_flavor_tag}.  In the case of a neutral-current interaction, the neutrino is not converted and merely transfers its energy and momentum into a target body. Neutral-current interactions occur through the sharing of a $Z^0$ boson. Due to the subtle nature of interactions mediated by the $Z^0$ boson, neutrino flavor signature is not easily evident. The NOvA near detector can detect and classify this weak interaction inter-mediated by the exchange of vector bosons $W^{\pm}$ and $Z^0$.

\begin{figure}[H]
\centering
\begin{subfigure}{.3\textwidth}
\centering
\includegraphics[width = \textwidth]{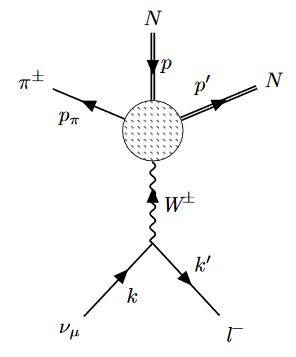}
\caption{\label{fig:Charged-current interaction}}
\end{subfigure}
\begin{subfigure}{0.3\textwidth}
\centering
\includegraphics[width = 0.9\textwidth]{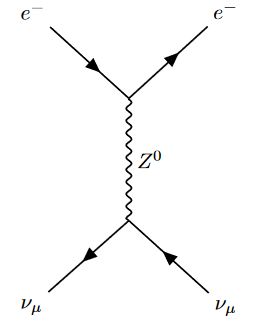}
\caption{\label{fig:Neutral-current interaction}}
\end{subfigure}
\caption[Feynman diagram of $\neutrino_\mu$ interactions.]{\label{fig:nu_mu-interactions} (a) Charged-current interaction. (b) Neutral-current interaction.}
\end{figure}
\addtocontents{lof}{\vspace{\normalbaselineskip}}

\setlength{\parindent}{10ex}
The NOvA experiment is a liquid-scintillator based tracking detector. Liquid scintillator neutrino detectors consist of hollow (typically plastic) tubes filled with organic material. Particle interactions within the scintillator result in scintillation and Cherenkov light emitted from within the medium. Given the specified cross-sectional area of the detector, liquid scintillators are primarily chosen for its cost-effectiveness and time resolution. When arranged strategically with scintillator signal attenuation considered, these detectors have good angular and position resolution as well. An added benefit to liquid scintillators is their low energy thresholds capable of detecting a wider array of neutrino energy interactions.
\newpage

As previously mentioned, the NOvA experiment is a tracking-based experiment that reconstructs the path of neutrinos and associated particles produced in charged and neutral-charged interactions. The greater the energy of the neutrino, the greater the path these particles travel in both types of interactions. Thus, tracking-based detectors must be very large in volume. Tracking detectors perform well at identifying particles form high energy neutrino interactions as longer tracks are easier to reconstruct. They are also well suited at identifying electron or muon interaction solely off the type of tracks or showers generated. Due to the event reconstruction capabilities of tracking detectors, they are well suited for distinguishing different particles (pions, muons, electrons, etc..) within a single event. These tracking detectors, however, rely on a high energy neutrino source beam -- enter Fermilab’s NuMI (Neutrinos at the Main Injector). 

Event reconstruction is based on an understanding of cross-sections. Described in a probabilistic sense of the type of interaction two particles will undergo under certain conditions, cross sections are essential in all neutrino experiments. The NOvA experiment studies neutrino cross sections in a variety of energy flux ranges with each type of scattering recorded\cite{OShegOshinowo}. The most abundant neutrino interaction is the charge-current quasi-elastic (QE) scattering with an energy flux spectrum less than 1 GeV\cite{behera2017event}. Neutrino (antineutrino) oscillation studies are the main point of interest in quasi-elastic scattering due to the recognizable signatures of all mediator particles involved. 

The next energy region within the NOvA experiment produces a resonance state (RES) which is attributed to a neutrino energy spectrum approximately between 1 GeV and 3 GeV\cite{behera2017event}. In the process of a RES interaction, a target nucleus produces pions in the final state. In this interaction, pions generated in the RES process may go on to interact with other nucleons of the target material or be absorbed, resulting in fake events. 

Finally, the third neutrino energy region of interest results in deep-inelastic (DIS) scattering. As neutrino (antineutrino) energies cross the threshold of 5 GeV, the DIS becomes the dominant interaction\cite{behera2017event}. In DIS, a highly excited neutrino scatters a quark within the target nucleons generating a lepton and multiple associated hadrons. With hadrons being the main product of collision, DIS is an important tool in the experimental world of particle physics studying hadronic properties. The Feynman diagrams below represent examples of each cross-section discussed\cite{Kutnink2012DetectorRC}.

\begin{figure}[H]
\centering
\begin{subfigure}{.3\textwidth}
\centering
\includegraphics[width = \textwidth]{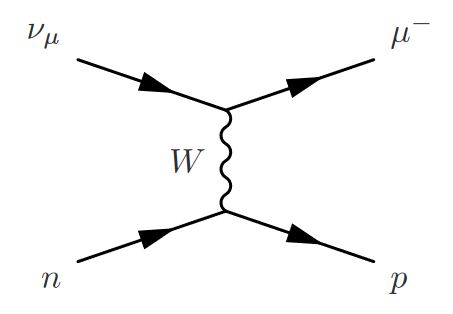}
\caption{\label{fig:QE}}
\end{subfigure}
\begin{subfigure}{0.3\textwidth}
\centering
\includegraphics[width = \textwidth]{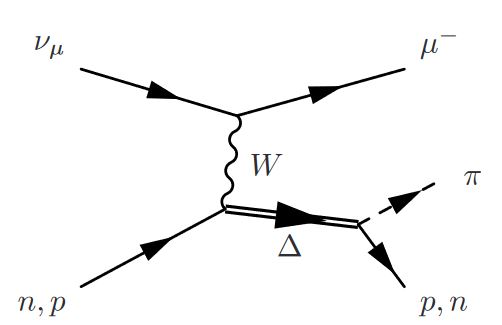}
\caption{\label{fig:RES}}
\end{subfigure}
\begin{subfigure}{0.3\textwidth}
\centering
\includegraphics[width = \textwidth]{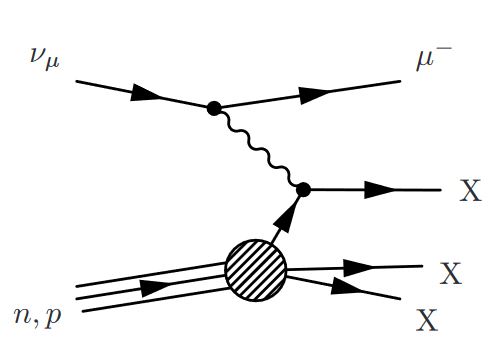}
\caption{\label{fig:DIS}}
\end{subfigure}
\caption[Feynman diagrams of Quasi-Elastic (QE), Resonance State (RES), and Deep In-elastic Scattering (DIS).]{\label{fig:scattering} Feynman diagrams illustrating (a) Quasi-Elastic (QE) (b) Resonance State (RES) (c) Deep In-elastic Scattering (DIS).}
\end{figure}
\addtocontents{lof}{\vspace{\normalbaselineskip}}

The cross-sections depicted above for each the quasi-elastic, resonance state production, and deep in-elastic scattering can be illustrated through respective CVN pixel map interactions. Recall the CVN pixel map is a single neutrino interaction extracted from the near detector global view. The neutrino energy, increasing from the left image to the right image is characteristic of each interaction mode.

\begin{figure}[H]
\centering
\begin{subfigure}{.3\textwidth}
\centering
\includegraphics[width = \textwidth]{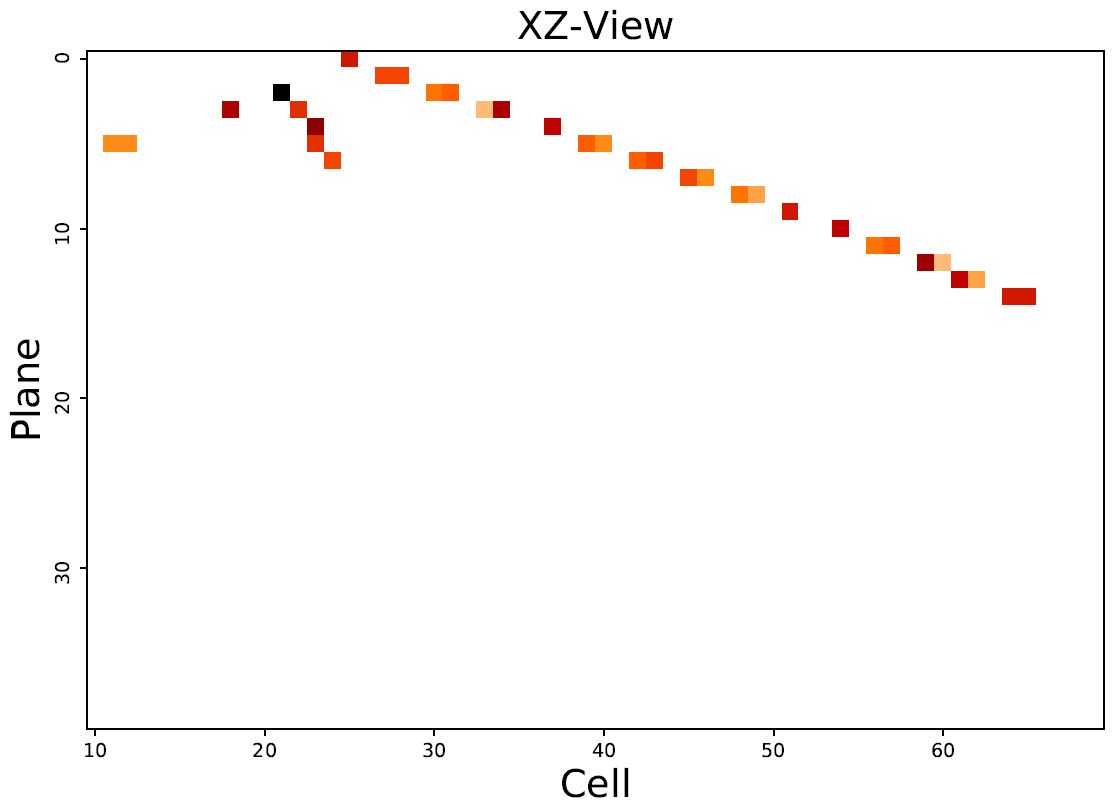}
\caption{\label{fig:cvnpmQE}}
\end{subfigure}
\begin{subfigure}{0.3\textwidth}
\centering
\includegraphics[width = \textwidth]{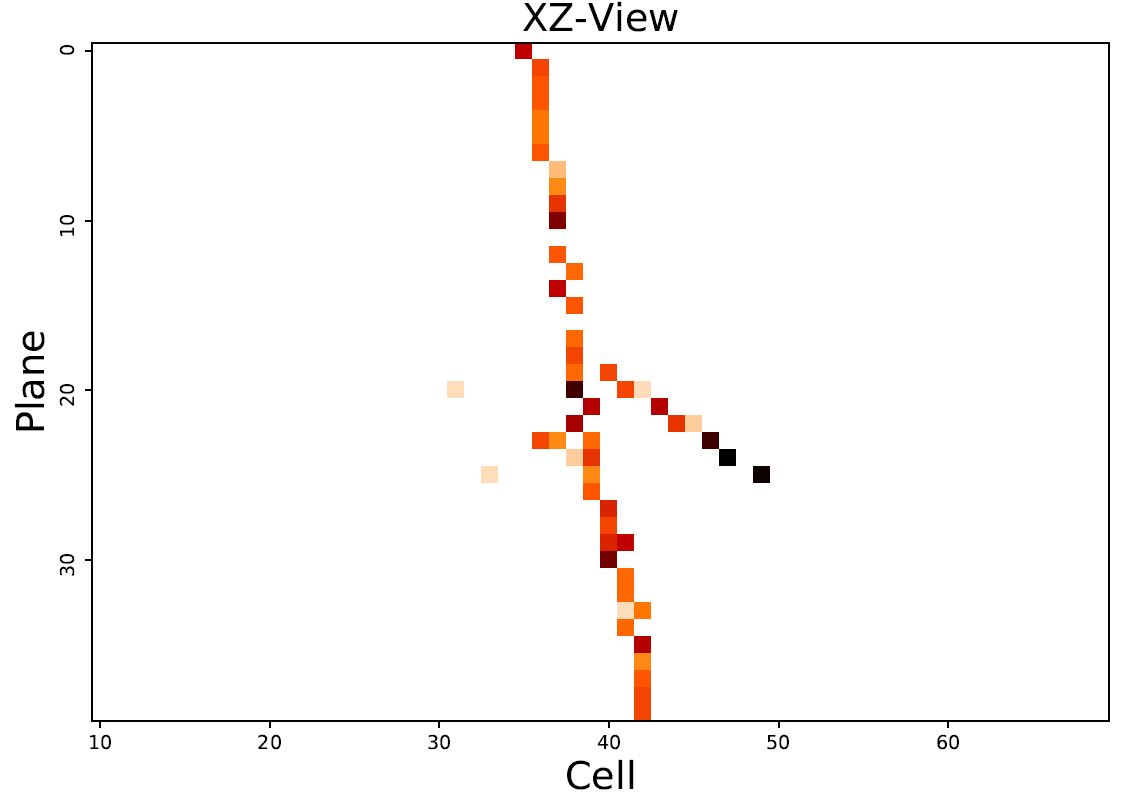}
\caption{\label{fig:cvnpmRES}}
\end{subfigure}
\begin{subfigure}{0.3\textwidth}
\centering
\includegraphics[width = \textwidth]{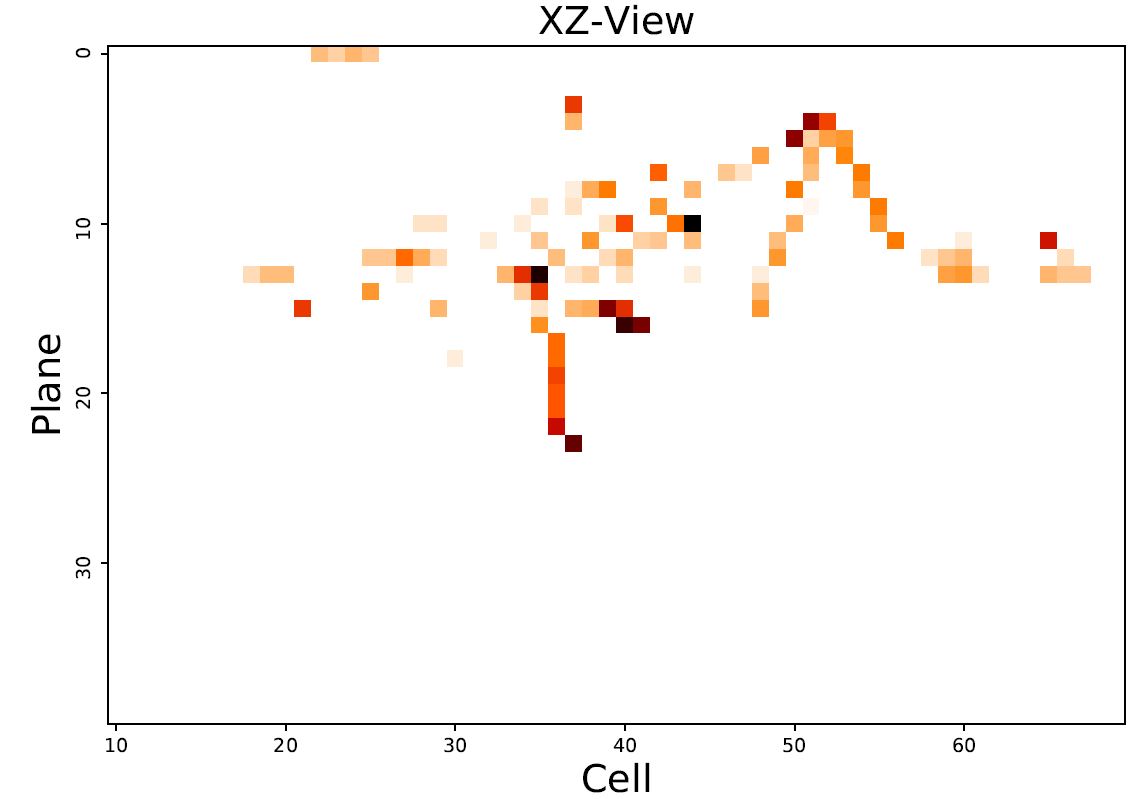}
\caption{\label{fig:cvnpmDIS}}
\end{subfigure}
\caption[CVN pixel map XZ view of Quasi-Elastic (QE), Resonance State (RES), and Deep In-elastic Scattering (DIS).]{\label{fig:cvnpmscattering} CVN pixel map XZ view of Quasi-Elastic (QE), Resonance State (RES), and Deep In-elastic Scattering (DIS).}
\end{figure}
\addtocontents{lof}{\vspace{\normalbaselineskip}}

\newpage
\setlength{\parindent}{10ex}
NuMI generates a high energy flux of neutrinos used by both the NOvA near and far detectors, as well as other Fermilab experiments. Protons strike a graphite target producing pions and kaons which are focused and directed using two electromagnetic horns. The pions are steered down a decay pipe that allows the pions to decay into muons and muon neutrinos. Any hadrons and muons are filtered from this beam through absorbers and layers of rock downstream. The final product is a high energy beam of muon neutrinos. The NuMI has two beam modes: forward horn current (FHC) and reversed horn current (RHC). The FHC mode sets the horn polarity to concentrate positively charged secondary beam particles, resulting in primarily a muon neutrino beam. When the horn polarity is reversed, in the case of the RHC mode, the beam is primarily muon antineutrinos at the 3GeV energy spectrum, with a good portion ($\approx$17\% of events) of muon neutrinos at the higher energy spectrum\cite{escapingeventsfd}. Below is a simplified schematic highlighting the major components of the NuMI assembly. The following sections will describe the near detector, its major components, and event data handling critical in understanding the reconstruction process.

\begin{figure}[H]
  \centering
    \includegraphics[width=6in]{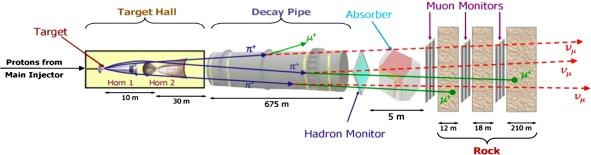}
    \caption[Component schematic diagram of NuMI beam line]{\label{fig:numi_schem_dia} Component schematic diagram of NuMI beam line}
  \end{figure}
\addtocontents{lof}{\vspace{\normalbaselineskip}}

\newpage
\subsection{NOvA Near Detector}

\setlength{\parindent}{10ex}
The NOvA near detector is located approximately 900 meters away from the Neutrino Main Injector target and 14.6 milliradians off-axis from the NuMI beam\cite{numi_beam}. This allows the near detector to receive a very narrow spectrum beam at the oscillation maxima. The approximately 300 metric ton NOvA near detector is sized at 4.2 m x 4.2 m x 14.6 m and sits in an excavated cavern approximately 100 meters underground to mitigate cosmic ray interactions. The near detector is built of prefabricated and assembled scintillator arrays arranged in zones. From closest to furthest from the NuMI beam target, these zones include a veto zone against tracks entering the detector, a target volume zone, a shower containment zone, and finally a muon ranger zone\cite{nova_tech_des}.

\begin{figure}[H]
  \centering
    \includegraphics[width=4.5in]{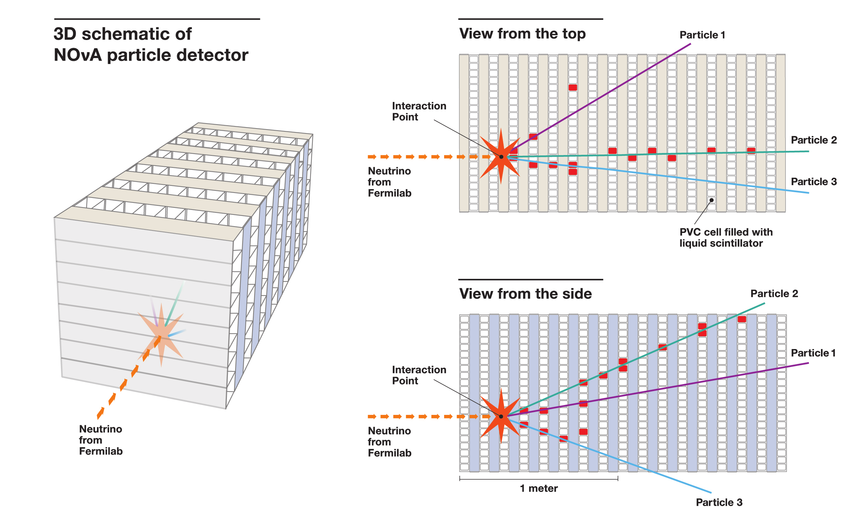}
    \caption[Schematic of NOvA particle detector]{\label{fig:nova_detector} Cell and plane layout schematic of NOvA particle detector}
  \end{figure}
\addtocontents{lof}{\vspace{\normalbaselineskip}}

At a basic level, the near detector is an assembly of modules comprised of PVC extrusions. To provide three-dimensional position data, each set of cells that form a plane are located orthogonal relative to the neighboring planes. These PVC extrusions are classified as either cells or planes depending on the viewing orientation. A PVC cell would be looking through the extrusion from one end, whilst a plane would be a cross-sectional view along the longitudinal axis. Altogether, there will be 199 planes of PVC liquid scintillators: 99 horizontal planes and 100 vertical cells. 

The effectiveness of the NOvA near detector heavily relies on the signal processing and data acquisition. At one end of the PVC cells, a front-end board (FEB) with wave signal attenuation fiber loop is installed. These FEBs are responsible for digitizing the scintillation light within the PVC extrusion and converting this signal into accurate position and intensity data. The mediator between the FEB and the wave-length shifting fiber is the avalanche photo-diode (APD). Another critical component to the experiment, the APD amplifies the signals within the PVC extrusion for the FEB to process. Finally, all data from the FEB is aggregated by a Data Concentrator Module (DCM) that transmit the digitized hit signals to a processing farm. Hits within the PVC extrusion contain spatial and time information, as well as a categorization of hit charge (ADC)\cite{nova_tech_des}.

\begin{figure}[H]
  \centering
    \includegraphics[width=6in]{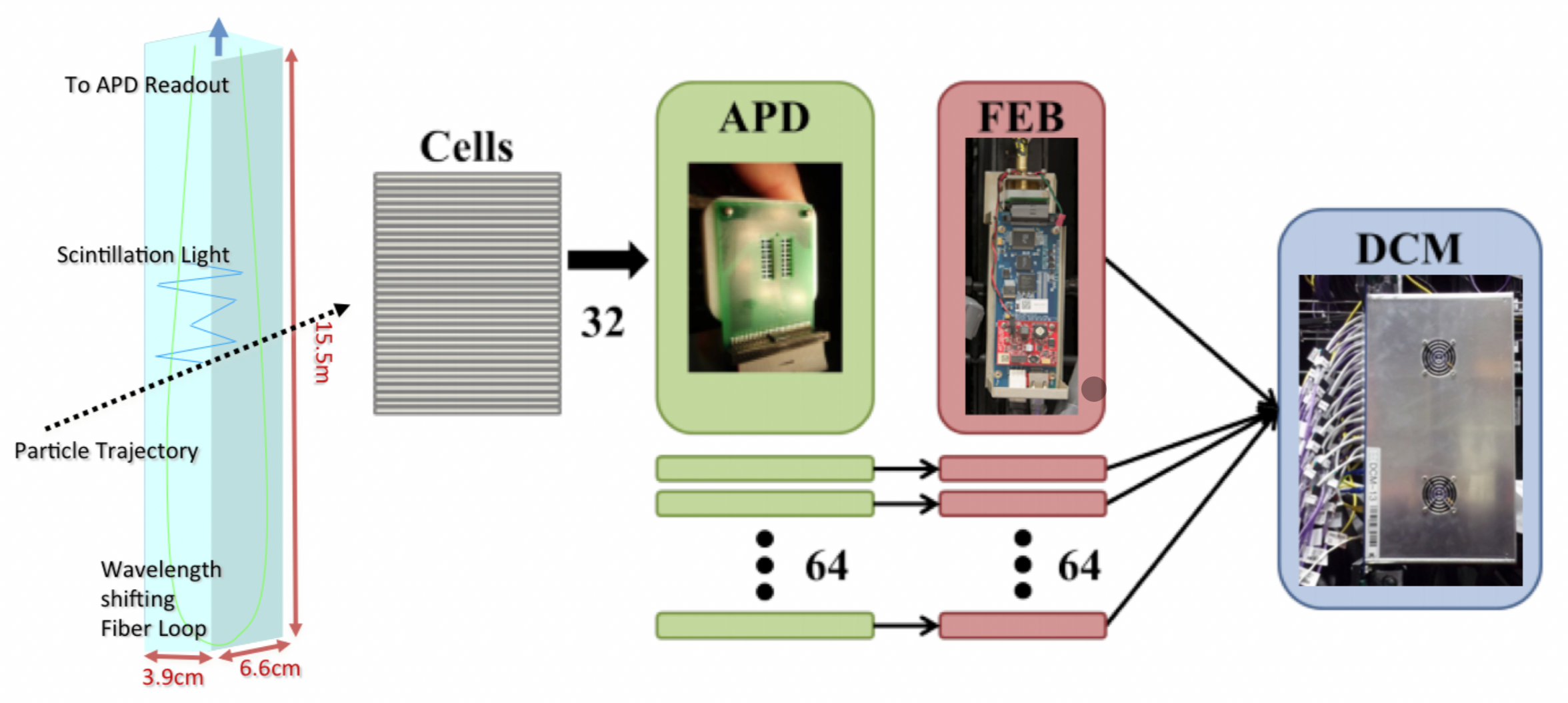}
    \caption[Scintillator hit handling]{\label{fig:nd_hit_proc} Flowchart of scintillator hit data handling within the NOvA system}
  \end{figure}
\addtocontents{lof}{\vspace{\normalbaselineskip}}

\newpage
\setlength{\parindent}{10ex}
The data acquisition (DAQ) system funnels the raw APD channel read outs and further processes this into a single stream of human-readable and archival information. The DAQ system caches the detector hit information until it is determined whether this data is worthwhile of storing or rejecting. The processing farm handles the information supplied by the DAQ system which performs complex quality assurance and testing before pushed to event building and data logging. The NOvA DAQ system has evolved to incorporate tiered dataset artdaq (root) framework. Researchers may access NOvA raw data using a data catalog to help narrow the scope of information by a variety of filter schemes. Over the years, NOvA reconstruction teams now offer different raw data formats with associated metadata. This work will be using simulated Monte-Carlo near detector data validated against real-world data. The application of this work is not limited by the use of this form of data.

\subsection{Event Tracking and Vertexing}

\setlength{\parindent}{10ex}
Recall a hit describes the information from a single triggered channel whilst an event is an aggregate of hits collected during a user defined time window. Hits can be classified as signal or noise. Signal hits occur over a set of signal channels. Hit density is computed by observing the number of hits within a sample region. These density hit values assist in event clustering algorithms that segregate signal hits and background noise. Metrics such as neighbor score, completeness, and purity based on neighboring hit density, proximity and energy support a data-driven approach to event reconstruction. Clustering algorithms support the first true step of event reconstruction known as slicing. Slices are the foundation for event vertexing and track finding. Next, a modified Hough transform is applied within the sliced hit area that seeds the slice to a global vertex in three-dimensional space. The seeded vertex found from the modified Hough transformation is used by a custom algorithm known as Elastic Arms that optimizes vertex candidates. Then, the vertex is used to seed a fuzzy k-means algorithm that generates prongs. Prongs are defined as a collection of hits associated to a single particle within a slice.  Finally, with the help of all these characteristics, a convolutional visual network is used to generate a pixel maps to characterize events and provide additional useful data for reconstruction purposes.  The flowchart below and following subsections elaborate on the steps involved\cite{CatanoMur:2018}.

\begin{figure}[H]
  \centering
    \includegraphics[width=6in]{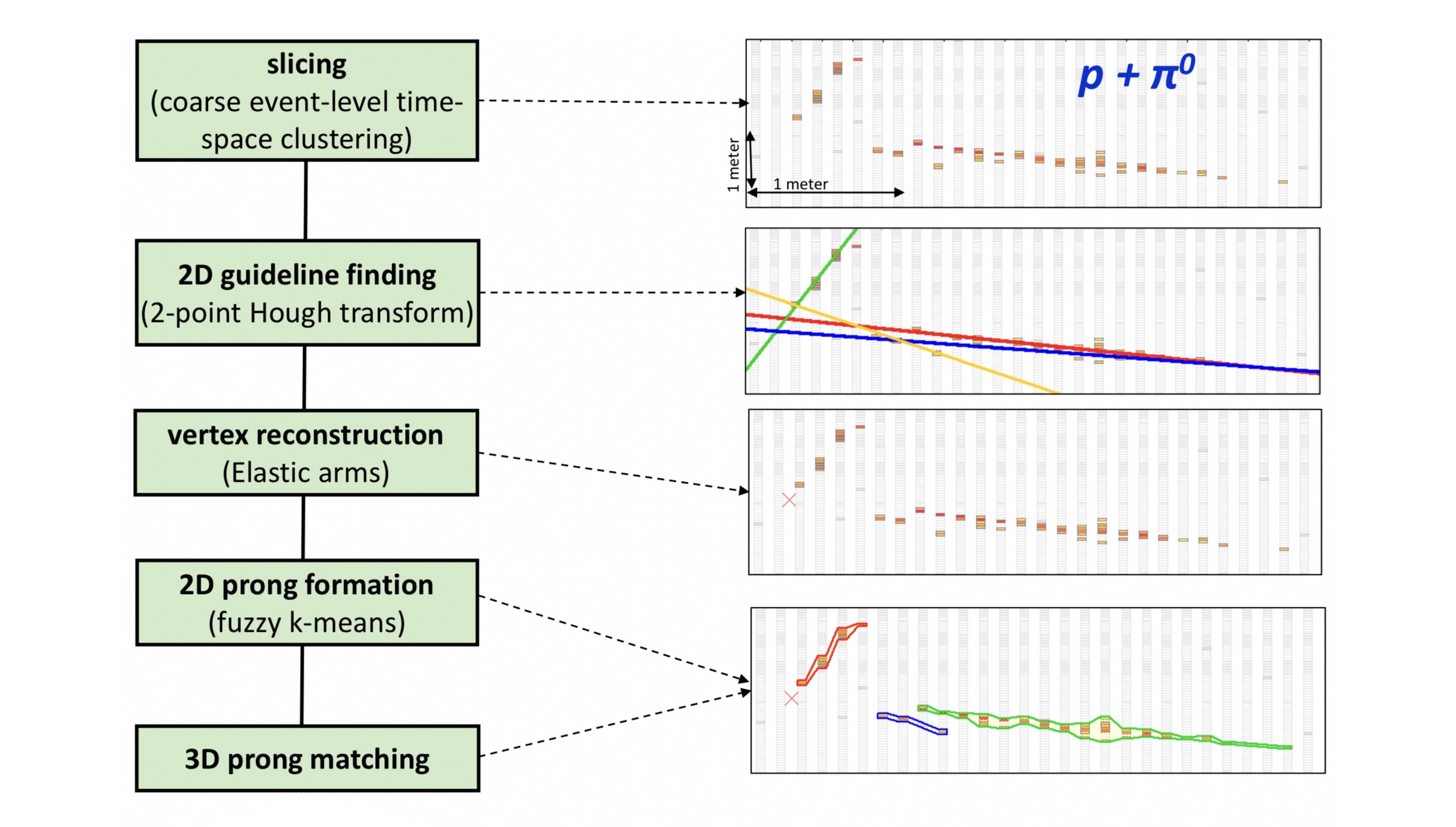}
    \caption[Event reconstruction algorithm process flowchart]{\label{fig:nd_reco_proc} Event reconstruction algorithm process flowchart}
  \end{figure}
\addtocontents{lof}{\vspace{\normalbaselineskip}}

\subsubsection{MultiHough Transformation}

Using a modified Hough transform algorithm, lines are identified in the next step after event slicing. This Multi-Hough transform utilizes an input of polar-coordinates ($\rho$, $\theta$) points characterizing a straight line through hits; where $\rho$ describes the perpendicular distance from the origin to this best fit line and $\theta$ the angle between x-axis and $\rho$\cite{behera2017tracking}. The algorithm fits these Hough defined lines within each independent view. From there, a Hough density map is generated with a set threshold and peak values favoring a line found to be most suitable for the hits in an area. The Hough algorithm is an iterative process that continuously generates new Hough density maps with new thresholds and peak values until no new peaks within the Hough space. The dominant Hough lines are presented with the intersections of these lines defining the primary vertex of this slice. The Multi-Hough algorithm has been optimized by ignoring hits that are distant from 2D tracks. Lines that would set 'fake' tracks are mitigated by optimizing the iterations on the threshold value. Another improvement includes the removal of multiple adjacent lines plotted against a set of nearby hits by removing the Hough line with the lesser number of hits.

\subsubsection{Elastic Arms}

The next reconstruction step after slicing and the Hough transform is running an 'elastic arm' algorithm  on the slice to determine the primary interaction point (vertex) of the neutrino\cite{behera2017tracking}. By seeding the predetermined the intersection of the Multi-Hough lines, the output of the elastic arm algorithm is a global 3D vertex point. The elastic arms algorithm is based on a deformable templates method that draws a straight line defined in three-dimensional Cartesian space by a polar and azimuthal angle ($x$,$y$,$z$,$\theta$,$\phi$). The algorithm finds parameter values that minimize the energy function below to describe the event topology.

\begin{equation}\label{eq:eaenergy}
\textit{E} = \sum_{\textit{i}=\textit{1}}^{N}\sum_{\textit{a}=\textit{1}}^{M}{\textit{V}_{ia}}{\textit{M}_{ia}}+\lambda\sum_{\textit{i}=\textit{1}}^{N} \left(\sum_{\textit{a}=\textit{1}}^{M}{\textit{V}_{ia}}-1 \right)^{2}+\frac{2}{\lambda_{\neutrino}}\sum_{\textit{a}=\textit{1}}^{M}D_{a}
\end{equation}

\noindent
where \textit{M} and \textit{N} are the total number of arm and hits in a slice, respectively. $M_{ia}$ is the perpendicular distance from hit \textit{i} to the projected arm in 2D space. $V_{ia}$ describes the probability a hit \textit{i} is tied to an arm \textit{a}. $D_{a}$ relates to the distance between the vertex and first hit within arm \textit{a}. Finally, $\lambda$ and $\lambda_{\neutrino}$ restrict penalty terms.

The fit of the hits within the arms is evaluated. If an 'arm' fails to find a hit within its trajectory, a penalty is applied to the energy optimization function. A penalty is applied if the vertex position is located too far from the initial hit within the slice. This is repeated for all vertices within the event. The success of the elastic arms algorithm heavily relies on the performance of the MultiHough algorithm. The final global 3D vertex of an event is used to evaluate the efficacy of both the MultiHough and elastic arms algorithms. 
\newpage

\subsubsection{Fuzzy K-Means}

The next important step in event reconstruction is the formation of prongs. Prongs are a bounded set of hits with a single start point. Prongs are associated to cell hits within a slice using a probabilistic fuzzy-k means (fuzzy-k) algorithm\cite{behera2017tracking}. A strength of the fuzzy k-means is that it allows for hits with a low probability of association to a slice to be ignored and treated as noise. The "fuzziness" is attributed to the fact that a hit can belong to more than one hit, within some local [non-normalized] probabilistic range. Prongs are formed in the XZ and YZ-views separately, and then the prongs are matched between the two views to create a complete 3D prong.

\setlength{\parindent}{10ex}
The fuzzy-k algorithm originates at the vertex provided by the elastic-arms algorithm. It utilizes the [energy] density of hits within the slice and the distance of the hits to the vertex to provide uncertainty values along the trajectory. Prong membership begins with an assumption that one prong defines the highest density cell hit area. Over each iteration, a prong center point is added to expand the prong path. In this process, prongs do not follow a straight path and their incremental angles along each prong center is subject update as uncertainty weights are calculated. Additional prongs may be added until all cell hits within a slice have $\geq$1\% association to a prong and no more prong seeds may be added.
Prongs are split with major spatial gaps that suggest a proximate pair of particles. Prongs are merged if cell hits association can be extended due to significant overlap.  
\newpage

\subsection{Convolutional Visual Network (CVN)}

The NOvA event reconstruction process has expanded over the past few years to include deep learning tools aimed at automating many of the classification and neutrino energy ($E_{\nu}$) estimation functions\cite{Baldi_2019}. The NOvA reconstruction group has made big strides in the machine learning field through the use of an image recognition convolutional neural network termed the convolutional visual network (CVN). The CVN is designed on the Caffe modifiable framework. This highly modular framework is built on a C++ library with Python bindings for training, testing, and modeling convolutional neural networks intended for use in multimedia research applications\cite{cvnid}.

\setlength{\parindent}{10ex}
In terms of event reconstruction and the observation of $\neutrino_{\mu}$ disappearance and $\neutrino_{e}$ appearance from neutrino oscillation, the NOvA group has developed the CVN particularly for the identification of neutrino events and any future work in application of machine learning\cite{nova_ml_ws}. The CVN architecture performs down-sampling and feature finding using pooling and convolutional layers, respectively to identify neutrino events\cite{aurisano_2016}. Slices, clustered energy dense cells within a specific range in time, are the CVN input feature maps. Using simulated events, the CVN is trained on 3.7 million neutrino and cosmic ray interactions within the NOvA far detectors. The figure below is of the YZ-view of a $\neutrino_{\mu}$ charged-current interaction and the 64 convolutional filters -- the output of the trained filters. Higher intensity cells or hits indicate greater activation which can be associated to a particular interaction after an inception module is applied. This pattern recognition is the fundamental idea that supports the CVN neutrino interaction classification model. This CVN model considers the XZ and YZ-views separately. Each view is subjected to independent down-sampling, convolution, and local response normalization (LRN). At the conclusion of handling, the two views are merged and input into the last stage inception module to latter be pooled. An output is generated with a final \textit{softmax} classification layer.

\begin{figure}[H]
  \centering
    \includegraphics[width=6in]{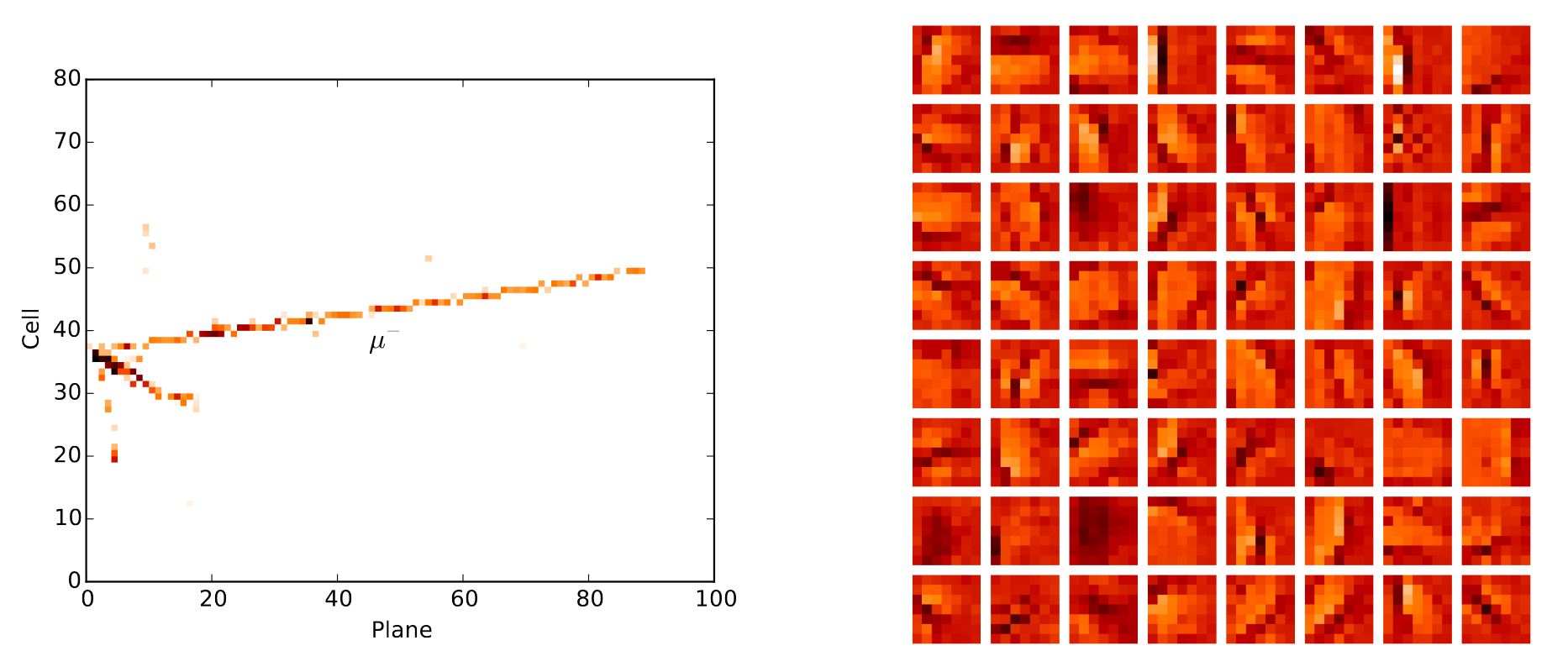}
    \caption[ND pixel map with convolutional filters and layers]{\label{fig:conv_filters_layers} Near Detector CVN pixel map with associated 64 convolutional filters and layers}
  \end{figure}
\addtocontents{lof}{\vspace{\normalbaselineskip}}

 This work heavily relies on the results from the CVN model by using the same pixel maps, interaction classification results, and other associated simulation data pulled from the NOvA art framework datasets. The NOvA experiment group has used hierarchical data format (HDF) to manage the CVN pixel maps and most of the associated data to each detector event. The NOvA art framework module generates the HDF files from a particle identification (PID) file. The PID files include higher order physics information, not just reconstruction relevant information useful for this work. Python manipulates HDF [version 5] files easily with NumPy and Pandas libraries. Using an architecture similar to Python's dictionaries and keys, extraction of data from HDF5 files is highly user friendly. The work in this paper heavily relies on the HDF5 data made available by the NOvA experiment group. To validate the results obtained from the deep learning model, simulation data must be used. It is important to note that the HDF5 files used here are pulled from the Near Detector (Reverse Horn Current) GENIE Monte Carlo dataset, 2019 mini-production edition. The availability of ND MC HDF5 simulation data is limited as tier conversion and calibration is still underway. Over the next few sections, further discussion will take place on the required data handling and pre-processing in preparation for the deep learning model.

\clearpage
\section*{CHAPTER III}
\vspace{0.25in}
\section{Improving Event Reconstruction}
\addtocontents{toc}{\vspace{\normalbaselineskip}}

\subsection{Limitations of Current Event Reconstruction Methods}

The aim of this work is to explore the possibility of improving event reconstruction through machine learning and to build the foundation for future efforts in secondary vertex finding. A review of the current tools and algorithms in use by the NOvA reconstruction teams show that though they deliver suitable results in event reconstruction efforts, they have their own limitations in aspects such as secondary track and vertex finding. The MultiHough algorithm was initially selected to be modified to determine secondary interaction tracks, but the constraints applied bias this method to detect the highest intensity and density hits for primary tracks and vertexing. 

\setlength{\parindent}{10ex}
As the MultiHough process relies on the iterative process of filtering out lower density hits, secondary events within a slice are at a disadvantage from being considered in Hough track plotting. The Hough algorithm is designed to consider these potential track indices as ‘fake’ tracks. The hit threshold cut off varies from one event to another determined by the HoughHisto distribution. In the case multiple tracks do get plotted adjacent to a set of hits, the tracks with the lesser hits are removed through a Hough track comparison matrix. Thus, eliminating possible secondary interaction tracks as well. A review of the MultiHough algorithm shows that it has been designed with the intent of determining the primary tracks and vertices. Modifications to a custom MultiHough algorithm for secondary track and vertex finding would ultimately lead to resurfacing new issues such as plotting of tracks with hits distant from the slice if the threshold point were modified.

The elastic arms and fuzzy-k algorithms rely heavily on the Hough transform conducted to seed tracks and prongs. The vertex found after the MultiHough and elastic arms processes has an average resolution of 11.6 cm from $\neutrino_{\mu}$ CC events and 28.8 cm NC events. Approximately 68\% of the CC and NC event vertices found through these methods are within 10 cm and 38 cm of the true vertex value, respectively. With interaction type yielding different results through traditional vertexing methods, it is important that this be kept in mind as the machine learning model analyzes the data.

In terms of secondary vertexing, it will not be possible during this work to predict the secondary vertex values through supervised machine learning as the true values are not provided in the Monte-Carlo simulation datasets. An attempt is made on proposing a kernel density estimation method to estimate tracks and candidates of secondary vertex within the CVN map. This would allow for future supervised machine learning methods to predict secondary vertexing. The figures below illustrate a few examples of the discrepancies found in the MultiHough, elastic arms, and fuzzy-k algorithms.

\bigskip
\begin{figure}[H]
  \centering
    \includegraphics[width=6in]{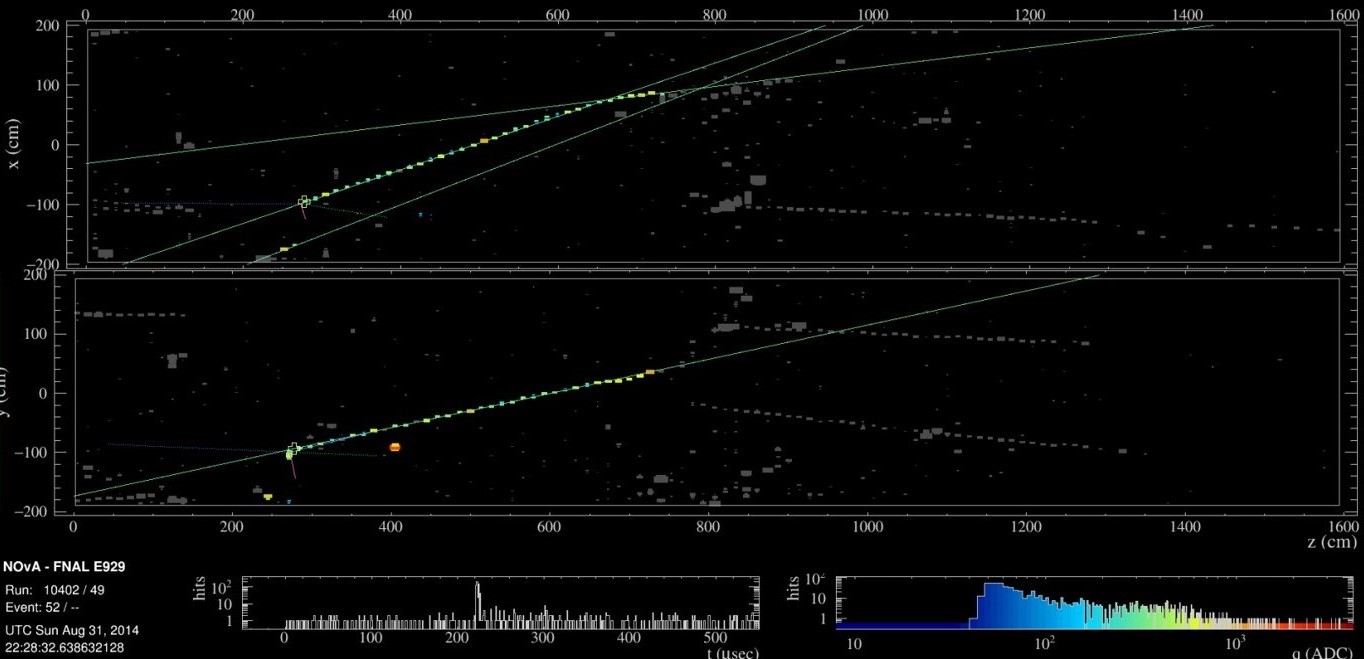}
    \caption[Sample event reconstruction discrepancy.]{\label{fig:sam_event_disc_1} This event depicts the reconstruction vertex (heavy yellow plus symbol) in the YZ-view to further down the detector than the true vertex. Edge background events were given Hough tracks.}
  \end{figure}
\addtocontents{lof}{\vspace{\normalbaselineskip}}

\newpage
\begin{figure}[H]
  \centering
    \includegraphics[width=6in]{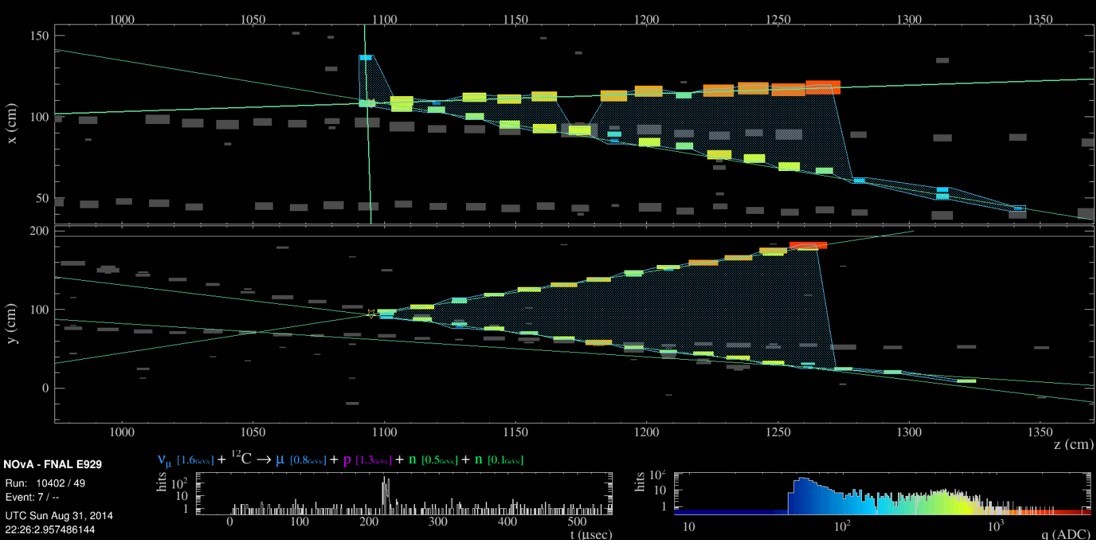}
    \caption[Sample event reconstruction discrepancy.]{\label{fig:sam_event_disc_2} In this event, the true primary vertex (heavy yellow star) is located forward of a hit along the z-axis. The Hough track plotted vertically in the XZ-view appears to have not been considered to correct for the Hough vertex position. The fuzzy-k algorithm appears to have also aggregated the full slice instead of the independent tracks.}
  \end{figure}
\addtocontents{lof}{\vspace{\normalbaselineskip}}

\begin{figure}[H]
  \centering
    \includegraphics[width=6in]{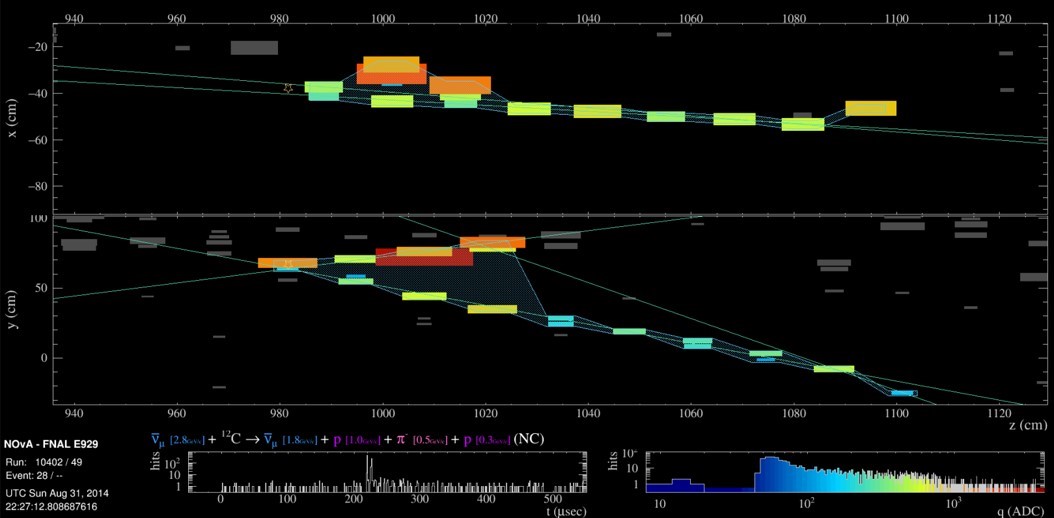}
    \caption[Sample event reconstruction discrepancy.]{\label{fig:sam_event_disc_3} The true primary vertex (heavy yellow star) and the Hough vertex (intersection of Hough lines) appear to agree in the YZ-view but not the XZ-view. A Hough track was plotted across the slice incorrectly. As well, a single prong was added to the full slice incorrectly. This is likely due to the density of hits in the XZ-view of the slice.}
  \end{figure}
\addtocontents{lof}{\vspace{\normalbaselineskip}}

\begin{figure}[H]
  \centering
    \includegraphics[width=6in]{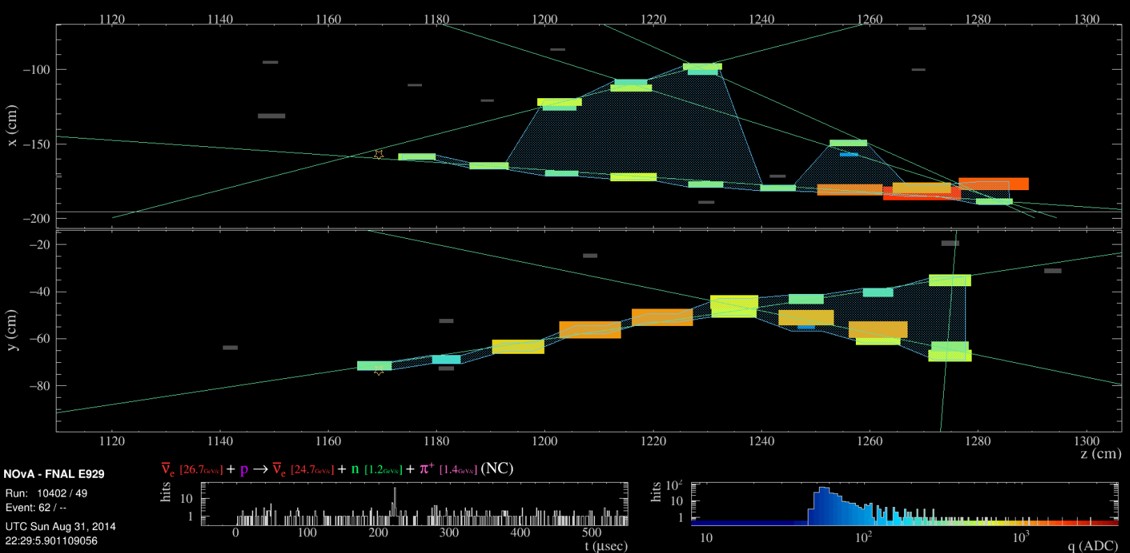}
    \caption[Sample event reconstruction discrepancy.]{\label{fig:sam_event_disc_4} In more active slices, the ability to determine reconstruction vertices becomes complex. The Hough vertex in the XZ-view appears to be close to the true vertex but the incorrectly placed Hough lines in the YZ-view due to provide a suitable reconstruction vertex candidate. Due to the Hough lines crossing through particles, prongs were also incorrectly placed to include the full slice.}
  \end{figure}
\addtocontents{lof}{\vspace{\normalbaselineskip}}

\begin{figure}[H]
  \centering
    \includegraphics[width=6in]{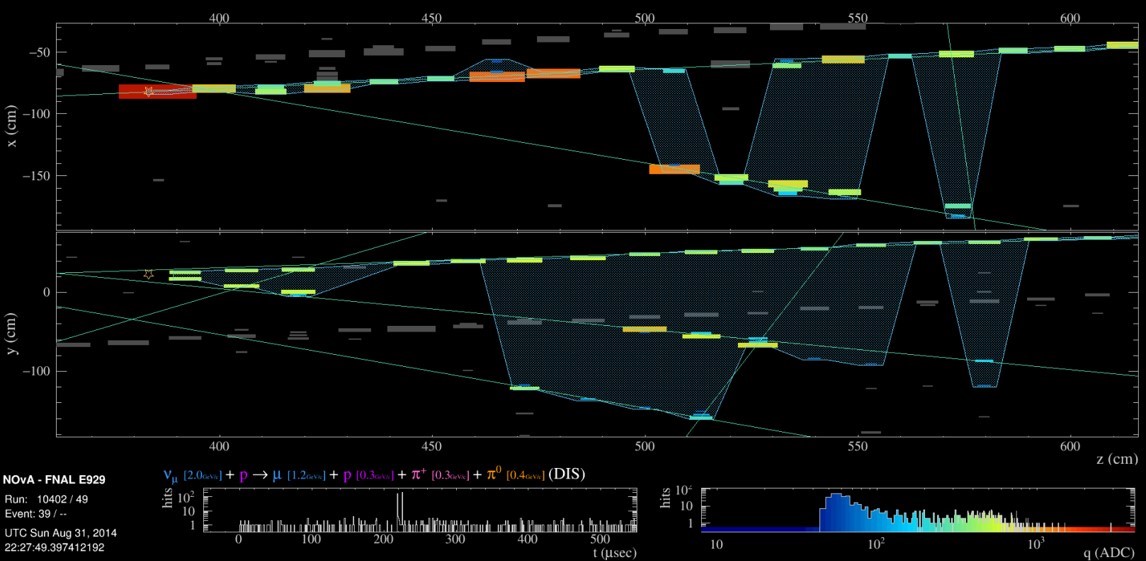}
    \caption[Sample event reconstruction discrepancy.]{\label{fig:sam_event_disc_5} In this event, the Hough primary vertex is placed in two different positions along the z-axis in both the XZ and the YZ-views. The true vertex (heavy yellow star) appears to be forward and aft of the Hough intersections. Also, it is important to note the Hough tracks and prongs crossing through the slice.}
  \end{figure}
\addtocontents{lof}{\vspace{\normalbaselineskip}}

\begin{figure}[H]
  \centering
    \includegraphics[width=6in]{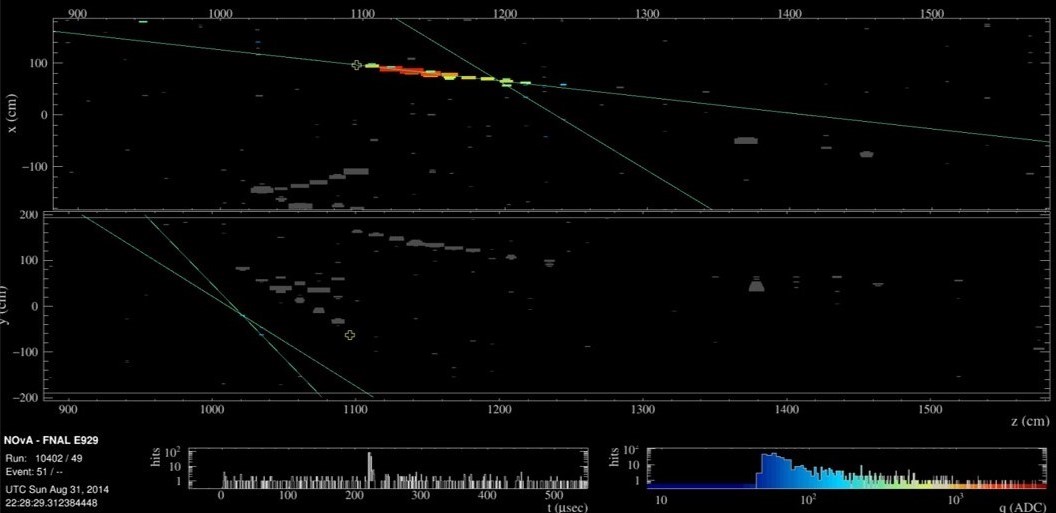}
    \caption[Sample event reconstruction discrepancy.]{\label{fig:sam_event_disc_6} In this example, the issue behind MultiHough working independently in each view is illustrated. One view, in this case XZ-view, must provide the full details necessary to suggest a candidate for a primary vertex. As the Hough lines provide no suitable candidates for a primary vertex, the elastic arms algorithm defaults to plotting a reconstruction primary vertex at the earliest point of the hit along the Hough track.}
  \end{figure}
\addtocontents{lof}{\vspace{\normalbaselineskip}}

\newpage
\subsection{Review of Deep Learning and Neural Networks}

\setlength{\parindent}{10ex}
With the volume of data made available to researchers continuing to grow and the exponential evolution of computing power, machine learning has found its way into more fields to solve more problems. Machine learning falls under the broad scope of artificial intelligence. A machine learning algorithm is capable of parsing and learning from data to then make informed decisions by applying what was learned. A subset of machine learning, deep learning utilizes its own computing processes to improve the informed decision process if an unsuitable outcome is met. A deep learning model is designed to constantly analyze the data logically to draw conclusions, similar to a human brain. This is done with the use of a layered algorithm structure called an artificial neural network, inspired by the biological neuron found in the human brain.

To better understand the complex structure of a neural network, a perceptron model is developed and discussed. Similar to a biological neuron with inputs (dendrites), an output (axon), and a processing point (nucleus), the perceptron is the building block of a supervised learning algorithm used in a variety of applications within an artificial neural network\cite{ml_intro}. Quantitatively, a perceptron is a function with adjustable parameters applied to the input(s). The adjustable parameters include weights multiplied to each input, and a bias term added to the inputs to account for nulls. The output of a perceptron is the aggregate of these weights and biases parsed through a user-specified activation function. Activation functions help set boundaries for the overall output value. A variety of activation functions exist, and it is critical that the appropriate activation function is selected for the output form of neural network. Examples of frequently used activation functions include sigmoid, rectified linear unit, Heaviside, Gaussian, and tanh functions. Considerations need to be made on the form of input data and range of output data, activation sensitivity, vanishing gradient, etc. The activation functions used in this work will be discussed later in detail.

\newpage
The main advantage of deep learning is that it does not require structured data; each network layer logically detects features or patterns of interest within the dataset. In the neural network model, a loss function is added in the feedback loop to update the weights and bias values. Simply put, a loss function is the measure of error between what the mode predicts and what the model output is valued\cite{ml_intro}. Loss functions could be regression based where a continuously evolving value is predicted, or classification based where the model predicts a specific category. Under these two broad categories, deep learning developers have provided an array of loss functions. Examples of regression loss functions include mean square error, mean absolute error, and root mean squared error. The loss functions explored in this work are regression based due to the nature of the outputs required (coordinate points within the near detector) and will be discussed further in the following sections. A schematic of a perceptron is provided to show the components involved in this feed forward process\cite{nueron_model}.

\begin{figure}[H]
  \centering
    \includegraphics[width=4.5in]{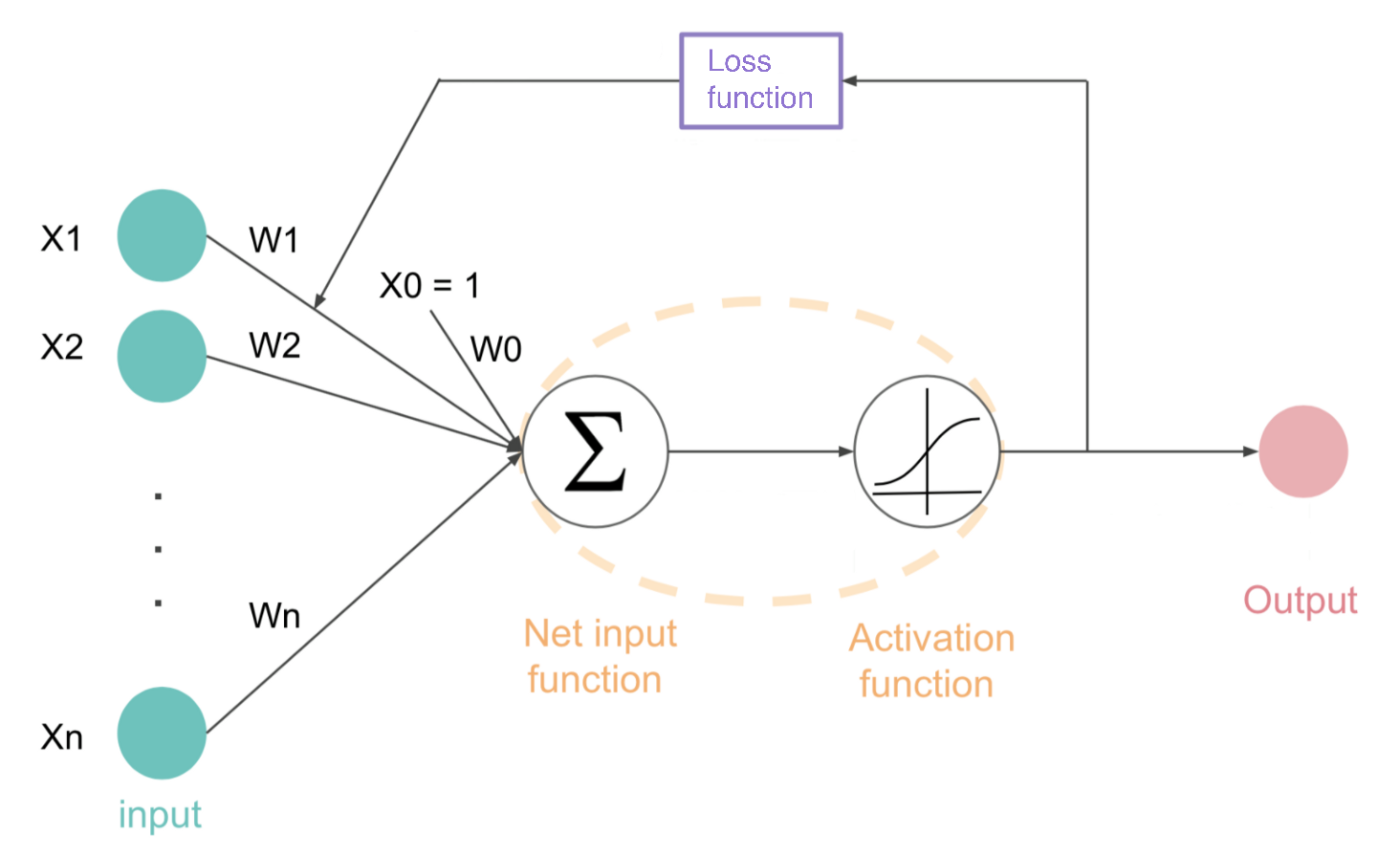}
    \caption[Neuron model]{\label{fig:neuron_model} Model of machine learning perceptron or neuron highlighting its fundamental components.}
  \end{figure}
\addtocontents{lof}{\vspace{\normalbaselineskip}}

\begin{equation}\label{eq:neuron_math_model}
\textit{y} = \sum_{\textit{i}=\textit{1}}^{n}{\textit{x}_{i}}{\textit{w}_{i}}+{\textit{b}_{i}}
\end{equation}

\noindent
where $y$ is the output, $x_i$ are the inputs, $w_i$ are the weight terms and $b_i$ are the biases terms for a tensor $x$ of $n$-dimensions.

\newpage
For complicated learning systems, this simple model of a perceptron can be expanded to include the inputs as a tensor (n-dimensional matrix) of information. By building a network of perceptrons, a multi-layer perceptron model is developed with the outputs of one vertical layer fed into the inputs of next adjacent layer. Layers may be full-connected, which means perceptrons [neurons] of one layer are connected to every neuron in the next layer. In a multi-layer model, the first layer is the input layer that directly receives the raw data. The final layer is the output layer, which can be more than one neuron in the case of multi-class classification model. The layers between the input and output layers are known as the hidden layers and are difficult to interpret due to the extensive inter-connectivity and distance from the input and output layers. A deep neural network is developed if two or more hidden layers exist.

\begin{figure}[H]
  \centering
    \includegraphics[width=5in]{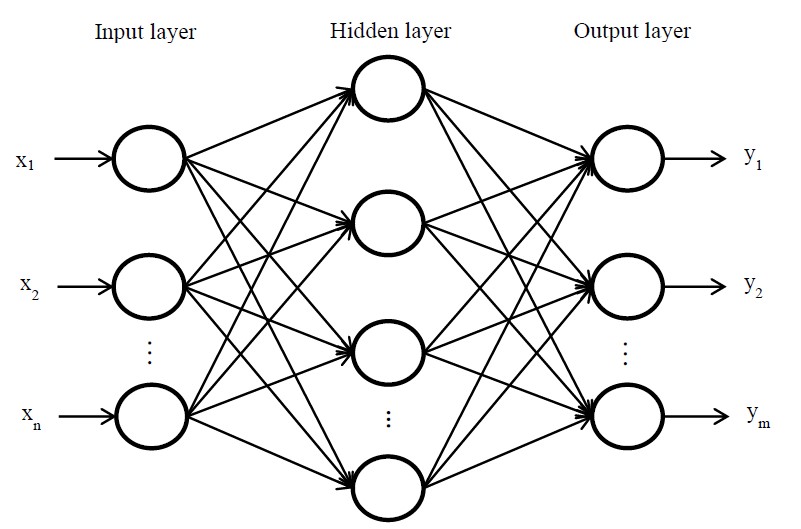}
    \caption[Representation of a neural network]{\label{fig:neuron_net_model} Representation of a neural network.}
  \end{figure}
\addtocontents{lof}{\vspace{\normalbaselineskip}}

\newpage
A convolutional neural network (CNN) is a specific architecture of neural networks that is extremely effective at working with image data. A common application of CNNs is taking input image data, processing it, and then classifying the images into categories. CNNs use multiple image filtering, pooling, and dense layers to achieve a user-desired learning objective. To better understand the CNN architecture, each type of layer and its function must first be explained.

The first layer of a CNN is the convolutional layer that extracts feature from the input image. When large number of parameters (for instance, number of image pixels) are input, a CNN applies multiple image filters to extract features that the layer can be trained on to change weight values. A CNN reduces parameters by focusing on local connectivity with neurons only connected to a specific set of adjacent local neurons that filter features. By reducing parameters, training time may be reduced compared to artificial neural networks.

In the case of a two-dimensional image, the neurons are locally connected based on the filter size and the stride each filter takes along the image. Multiple filters may be used depending on the complexity of the image and the features in consideration. The next consideration to be made is color channels. Color images may be thought of as three-dimensional tensors made up of red, green, and blue color channels. In this application of CNNs, one color channel is assigned to the hit intensity related to the energy deposited in each cell. Thus, a two-dimensional image is input into the CNN with the properties: pixel height, pixel width, and number of color channels.

Despite the local connectivity of neurons to filter image features, many parameters still exist that need to be processed. Pooling layers may help to further down sample or subsample the features. Pooling layers accept convolutional layers as input and have their own form of filters that apply a window with an associated stride extracting the max value in each kernel. A lot of information is removed this way; a 2 by 2 kernel with a stride of 2 removes 75\% of the input data. This may appear to hurt the model, but recall the goal is to reduce processing time and prevents units from  “co-adapting” or overfitting.

CNN architecture typically includes multiple dense layers within its hidden layers. Dense layers may be fully connected layers that perform the linear operations on the input data vector. Dense layers are added after the convolution and pooling layers have reduced the image parameters. The next CNN layer to describe is the flatten layer performs a tensor operation that converts that pooled feature map into a single column. This is done because the fully connected output layer can only accept data in this form. Other convolutional layers exist but the topics mentioned in this section cover the foundational understanding necessary to continue the event reconstruction work.

\begin{figure}[H]
  \centering
    \includegraphics[width=6in]{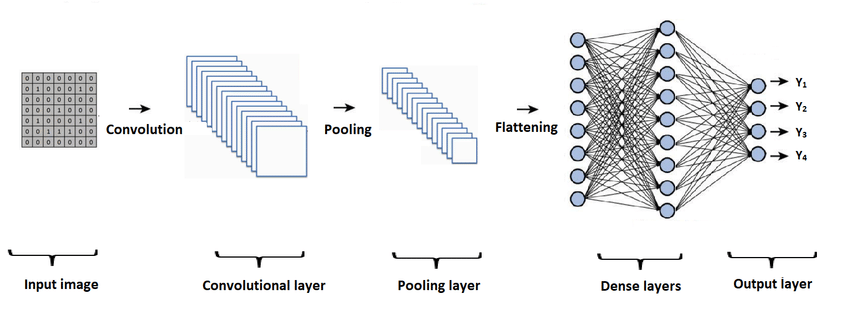}
    \caption[Representation of a convolutional neural network]{\label{fig:cnn_model}Representation of a convolutional neural network.}
  \end{figure}
\addtocontents{lof}{\vspace{\normalbaselineskip}}
\newpage

\subsection{Application of Deep Learning in the NOvA Experiment}

\setlength{\parindent}{10ex}
This research begins with the appropriate selection of data. The data used to build the training and validation sets is pulled from the standard NOvA simulation set. The simulated near detector dataset is prepared through a complex and computationally heavy process. This begins with generating neutrino beam flux using the FLUKA and FLUGG interface while cosmic rays are generated using CRY. The modeled neutrino interactions are propagated throughout the detector via GENIE which works in collaboration with GEANT4 to simulate energy deposits in the detector\cite{geant-sim}. In the last step, the energy deposits in the target material are parameterized in a program that converts the energy deposits into scintillation light, simulating the electronic responses in the APD. This will be the final format of the raw data. The complete NOvA data simulation process can be found in \cite{sim_chain}.

The raw data is validated, processed extensively in the ART framework, and then fed to the CVN feature map model before being published for use. The simulation data chosen must provide a variety of interactions and true data to validate any model predictions. The NOvA group has made available various formats for this data, but due to ease of use within Python, HDF5 was selected as the preferred format. As the transition is being made to HDF5 by the NOvA group, the available suitable simulation datasets are limited for the near detector. The simulation data used for this is the 2019 GENIE mini production version 5 reverse horn current near detector simulation which is a large and diverse set of events suitable for machine learning.

After the simulation dataset was chosen and transferred to local and cluster computers, initial event preprocessing and filtering must occur. The efficacy of a machine learning model heavily relies on the data provided; clean and simple images are easier for the model to handle and extract target features. A Python based HDF5 preprocessor was developed to reject $\neutrino_e$ cosmic rays from the CVN pixel maps, ignore interactions in the near detector veto zone, consider only $\neutrino_e$ and $\neutrino_{\mu}$ based interactions, and finally set a containment cut on prongs. As well, this HDF5 preprocessor extracted important attributes related to each event necessary for data parsing and model evaluation. This includes neutrino energy $E_\neutrino$, particle data group (PDG), true primary vertex values from the Monte-Carlo simulation, primary vertex values for each interaction computed through traditional reconstruction algorithms, and the classification of interactions into charged-current, neutral-charge, quantum elastic scattering, resonance state, and deep inelastic scattering. After the filtering of events, only 250,000 events were deemed appropriate for this study and the machine learning model.

The events were first split into a training and validation set at ratio of 70/30 percent respectively, and then again into interactions exclusively within the dimensions of the near detector. The CVN pixel map and associated data was handled in Python using NumPy and Pandas libraries. It is important to note that he CVN pixel map is not a total event capture within the near detector. In the raw HDF5 file, the CVN pixel map has the dimensions (2,100,80). Each CVN pixel map has both the XZ and YZ views stretching over 100 planes and 80 cells, centered over a slice. It was necessary to separate the XZ and YZ views and reshape the CVN pixel map to include a color channel. Each CVN pixel map set was then tied to a set of properties that are of interest to this study.

Once the data was parsed and ready for training, the deep learning model needed to be developed. A number of considerations had to be made: the number of convolutional layers, the number of dense layers which would affect the number of training parameters, the conditions of the training set, the activation function, the loss function, and of course how each view is to be handled. A regression-based approach convolutional neural network architecture was selected as the framework due to the type of images used and the output desired from the model. Ultimately, the network structure used in this work consists of two identical sub-networks for each view (XZ and YZ), merged to produce the final prediction. Figure \ref{fig:regcnn_model} depicts the neural network architecture. 

\newpage
The 2D-Convolutional layers for each view extract features at different locations producing the feature map which is down sampled by the MaxPool layers. The dense layers for each view develop the training parameters. Dense layers contain the neurons with the weight and biases values described in the previous section. In this model, each of the six dense layers per view has 256 neurons with rectified linear unit (ReLU) activation. This results in approximately 32 million trainable parameters for both views. The sub-networks are merged using a Tensorflow concatenation layer with a final dense layer with the number of classes [one] as the output. Through experimentation, adding more dense layers did not improve model performance evaluated by the loss versus epoch graph.

\begin{figure}[H]
  \centering
    \includegraphics[width=5.5in]{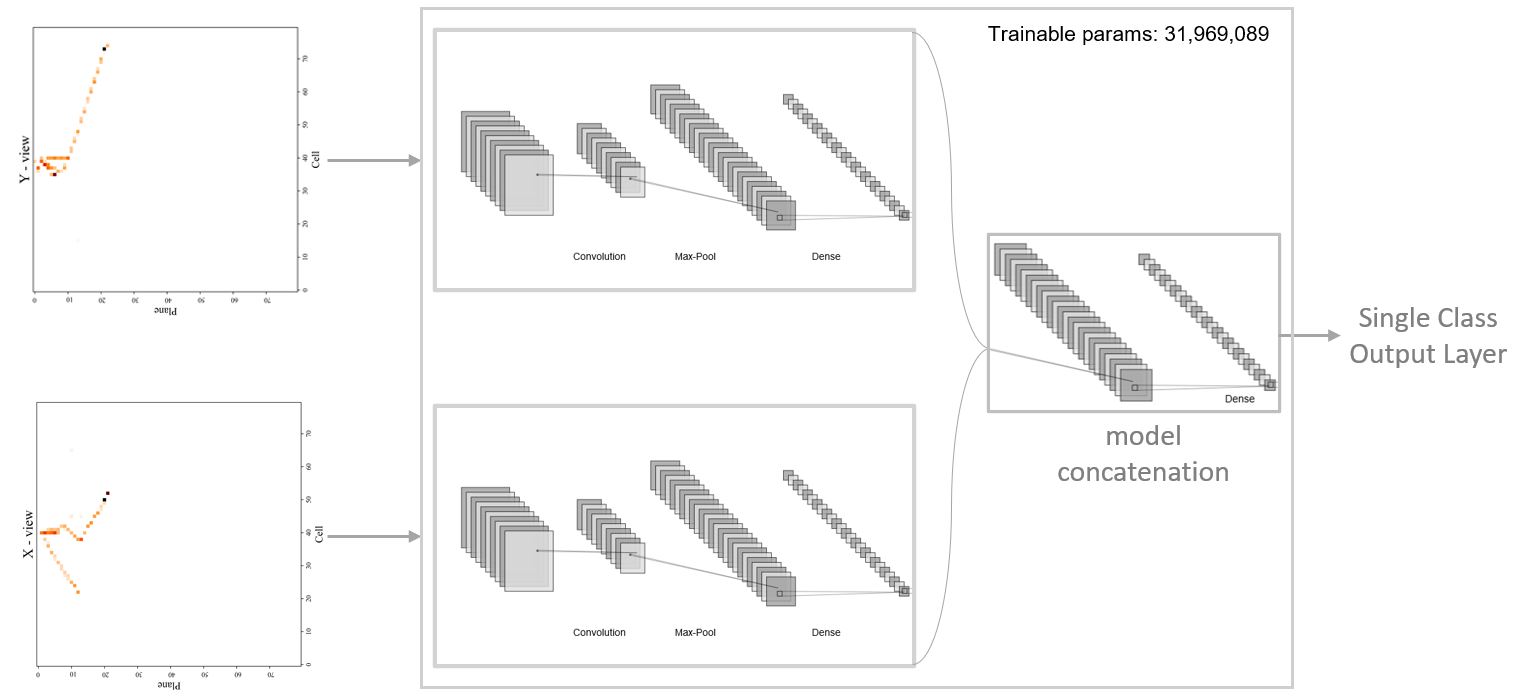}
    \caption[Two view sub-network regression based convolutional neural network]{\label{fig:regcnn_model}Two view sub-network regression based convolutional neural network.}
  \end{figure}
\addtocontents{lof}{\vspace{\normalbaselineskip}}
\newpage

Different regression loss functions were used, including mean average error (MAE), mean square error (MSE), and root mean squared error (RMSE). The loss function that yielded the best result was the logcosh function. The logcosh loss function takes the logarithm of the hyperbolic cosine of the prediction error and works similar to the mean square error\cite{intro_stat_learning}. The advantage of the logcosh function over the MSE is that it is not heavily affected by outliers in predictions. Due to the nature of the dataset and the span of potential vertices across the near detector, severely incorrect predictions should not heavily punish the model. This was learned from previous attempts at developing a deep learning model to predict slice vertices\cite{cvn_vtx_reco}. The logcosh function is still susceptible this issue if repeated off-target predictions are constantly made. The optimization method selected is the well-known Adam stochastic gradient descent algorithm. Adam is based on adaptive estimates and is computationally efficient\cite{ml_applied}. Given this dataset, Adam is more suitable than other optimizers for very large datasets that may be noisy.

\begin{equation}\label{eq:logcosh_loss_eq}
\textit{L}(\hat{{\textit{y}_{i}}},\textit{y}) = \sum_{\textit{i}=\textit{1}}^{n}{\textit{log}({\textit{cosh}}(\hat{{\textit{y}_{i}}}}-\textit{y}_{i}))
\end{equation}

\noindent
where $L$ is the loss function, $y_{i}$ is the true value, and $\hat{y}_{i}$ is the predicted value for the $i^{th}$ index.

\newpage
The model was trained with 70 percent of the 250,000 CVN pixel maps and true primary vertex positions. Limited initial training to validate the model was done on a local CPU, but the full dataset had to be done on a high-performance Graphics Processing Units (GPU). Wichita State University (WSU) has made available to students a high-performance computing (HPC) cluster for compute-intensive jobs. The HPC cluster at WSU, named BeoShock, has a total of four NVIDIA Tesla V100 16 GB Tensor GPU with 5,120 CUDA cores capable of 112 teraflops of floating-point calculations per second. The BeoShock HPC made it possible to experiment on different neural network models. This afforded the opportunity to gain new insight on different training sets by interaction type through iterative experimentation.

Three models were developed for each position in Cartesian space (x,y,z). Given the amount of data, the models were limited to train in 200 epochs to avoid overfitting. The training data spanned was not limited to interactions within the near detector. The validation data, however, was limited to interactions just within the near detector. This limitation was a criterion decided on as a point of interest to researchers. The primary vertex predictions made by the model were evaluated against true vertex values and traditional reconstruction vertex values. The results are discussed and analyzed in the next chapter.

\newpage
\subsection{Application of Kernel Density Estimation in the NOvA Experiment}
In this work, the KDE algorithm generates a plot along the z-axis independently in the XZ and YZ-views. The peak values of this KDE function strongly correlate to areas of clustered tracks and higher energy deposit regions. The vertex values may be offset from the peak values marginally but provide an approximate means to evaluate candidates through a supervised machine learning model. These peak values are then classified as either primary vertex or secondary vertex candidates depending on the number of tracks generated from that point. This is where the KDE function plotted from the planes and cells is useful. Using the information from this supporting algorithm, secondary vertex candidates may be graded higher or lower if the number of tracks propagating from this intersection is canonical. This performance metric may be tuned, for example secondary vertex candidates would need to have two or more tracks.

When future NOvA GENIE ND simulations offer true secondary vertex values, the candidate secondary vertex values may be used as the input array to the model, instead of the CVN pixel map. A sample result is plotted below as a proof-of-concept and limited due to the scope of this work. A similar hybrid approach is currently being reviewed by the Large Hadron Collider reconstruction research group as a means of finding secondary interaction vertices\cite{Schulz:2013}.

\begin{figure}[H]
  \centering
    \includegraphics[width=3in]{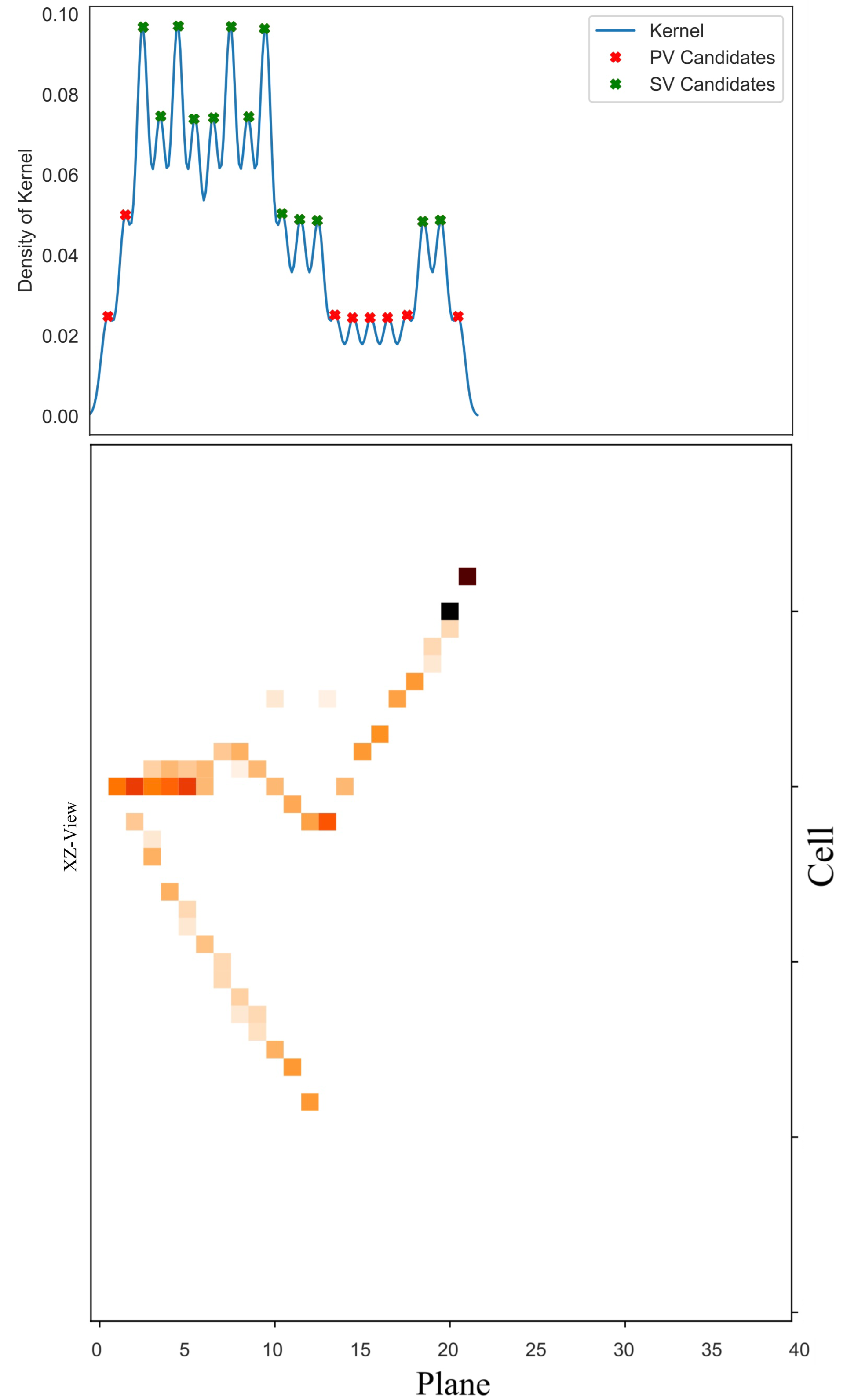}
    \caption[Sample application of Kernel Density Estimate function along the Z-axis in the XZ-view.]{\label{fig:kde-cell}Sample application of Kernel Density Estimate function along the Z-axis in the XZ-view. This generates candidates for primary and secondary vertex candidates from the density of hits weighted by energy deposited.}
  \end{figure}
\addtocontents{lof}{\vspace{\normalbaselineskip}}
\newpage

\bigskip

\clearpage
\section*{CHAPTER IV}
\vspace{0.25in}
\section{Results and Analysis}
\addtocontents{toc}{\vspace{\normalbaselineskip}}

\subsection{Model Prediction Results of Event Primary Vertex}

After validation runs of the custom regression-based CNN model on a small scale, the production version was run on the BeoShock HPC with the 175,000 events. The final data processing and training run for each coordinate position within the NOvA near detector was completed in approximately eight hours on the HPC. The model was validated against the 75,000 events split from the total number of events available. The validation events were classified with the characteristics: QES, RES, DIS, CC, NC, and a Boolean of whether the event was within the detector bounds. The objective was to test the efficacy of a regression-based CNN to predict primary vertex values from the CVN pixel maps. The results yielded were compared to both the true and current primary vertexing methods.

An analysis of the density plots shows that the application of machine learning, even for event vertexing, has a promising future in high energy physics. Charged-current events bound within the detector appeared to follow the true vertex values closer compared to neutral-charge events. Future work in the application of machine learning would need to dive into the most suitable training events for the mode of interaction. For example, training on CC events would likely lead to more accurate results on validation sets. In terms of secondary vertexing, due to the complexity and nature of secondary interactions, supervised learning of secondary vertex true values may not yield successful results. A hybrid approach of a secondary vertex candidate generator through a KDE algorithm and a deep learning model is likely the best course of action in future work. 

This work developed a CNN deep learning model that directly accepts neutrino event pixel map inputs for primary vertex prediction. This is an improvement on past CNN-based NOvA work and a foundation for future CNN based work aimed at predicting regression related problems. Enhanced particle tracking and measurement through improved primary vertexing would help with better estimating particle momenta and observing neutrino oscillations. Improved vertexing methods through this work may lead to the application of a regression approach CNN in other reconstruction methods such as the prediction of particle energies from the track length or the ability to detect secondary vertexing\cite{lhc_sv_ppt}. The effectiveness and performance of this primary vertexing CNN is shown through the prediction of various interaction modes and charges. This CNN has also shown smaller errors compared to traditional vertexing methods when the model is trained strategically.

\begin{figure}[H]
  \centering
    \includegraphics[width=6.5in]{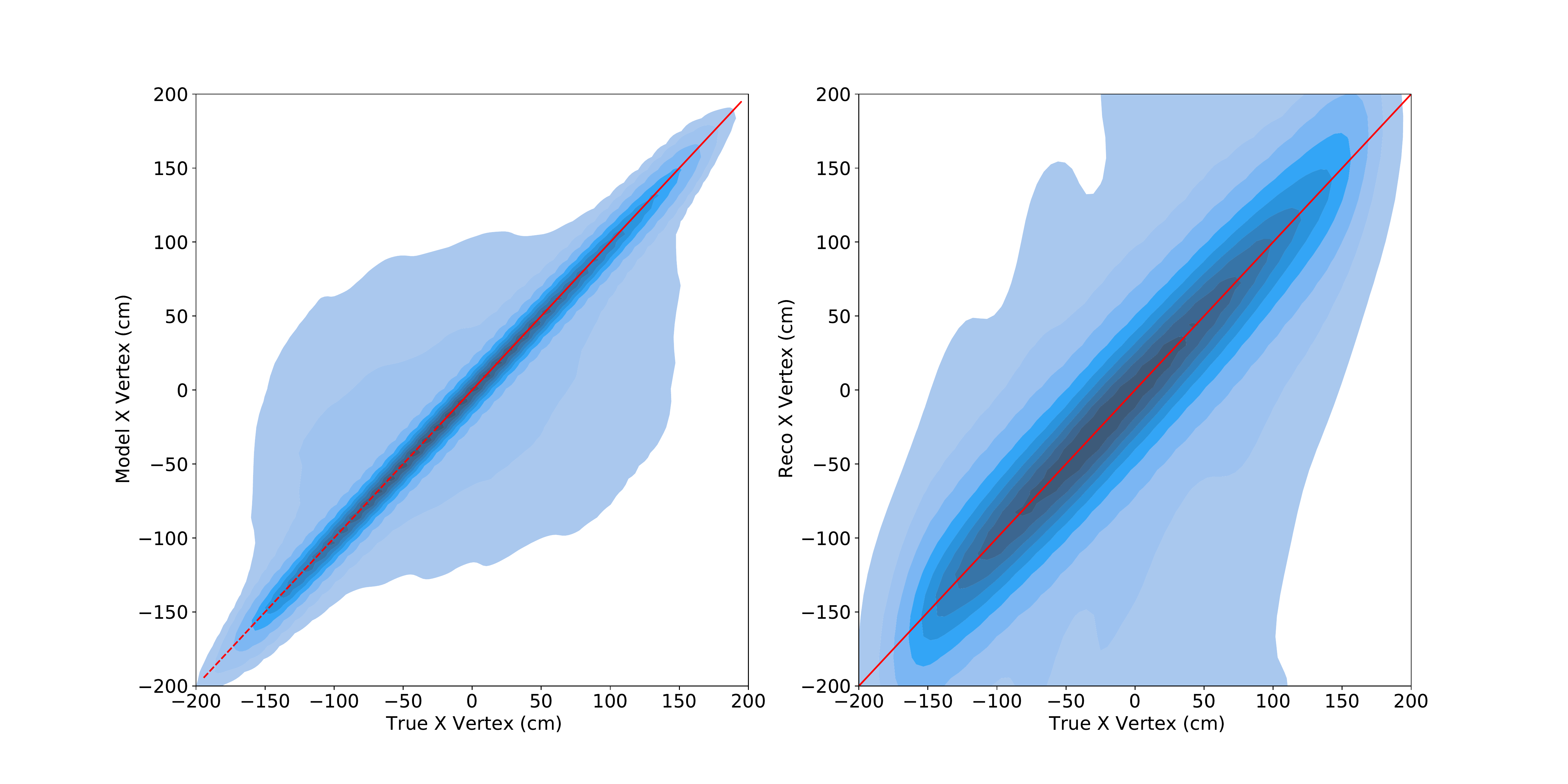}
    \caption[RegCNN model predictions and reconstruction vs true values along x-axis.]{\label{fig:x-full-modelpred-true-vs-reco-kde}RegCNN model predictions and reconstruction vs true values along x-axis.}
  \end{figure}
\addtocontents{lof}{\vspace{\normalbaselineskip}}
\bigskip

\begin{figure}[H]
  \centering
    \includegraphics[width=6.5in]{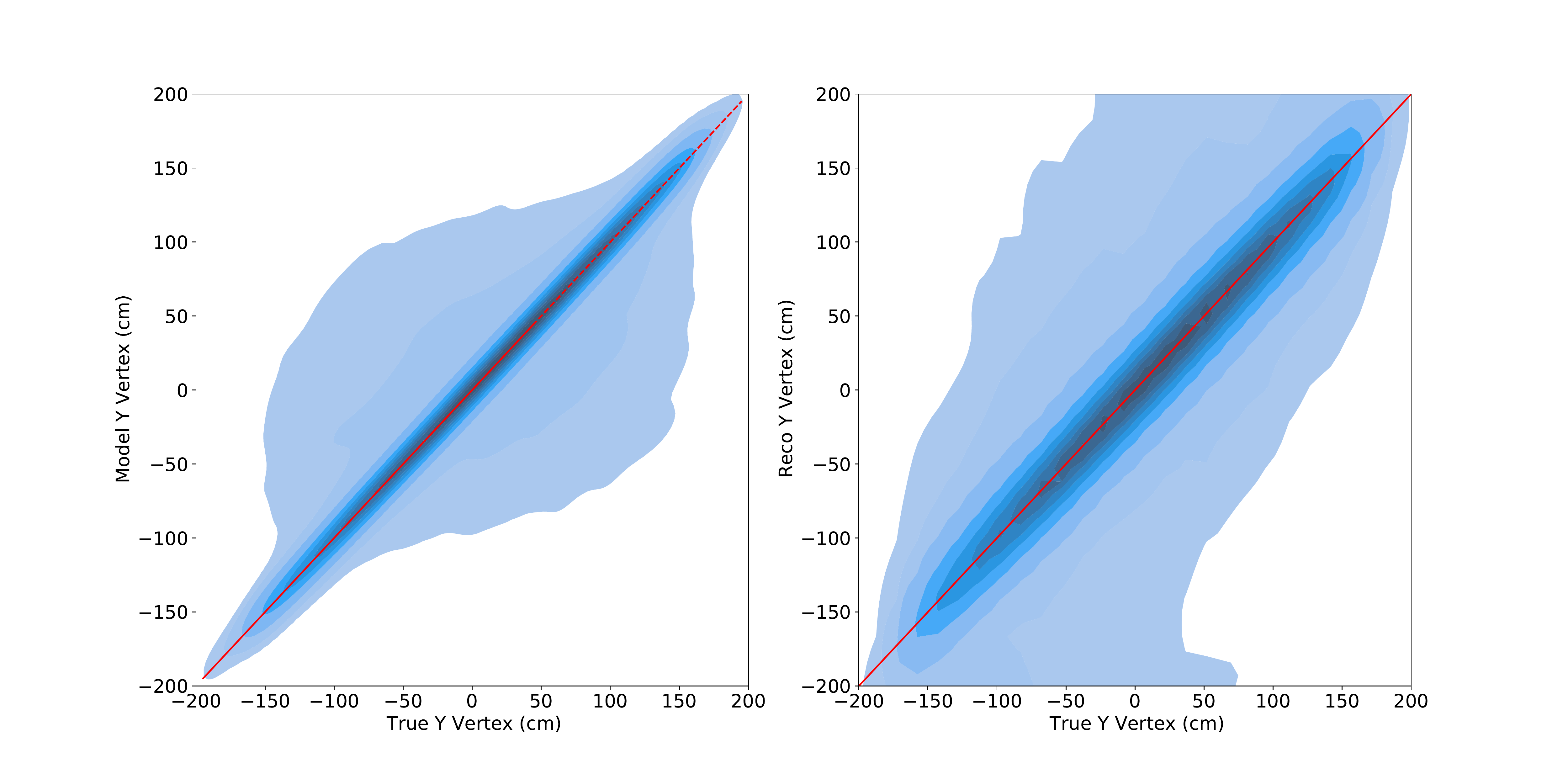}
    \caption[RegCNN model predictions and reconstruction vs true values along y-axis.]{\label{fig:y-full-modelpred-true-vs-reco-kde}RegCNN model predictions and reconstruction vs true values along y-axis.}
  \end{figure}
\addtocontents{lof}{\vspace{\normalbaselineskip}}
\bigskip

\begin{figure}[H]
  \centering
    \includegraphics[width=6.5in]{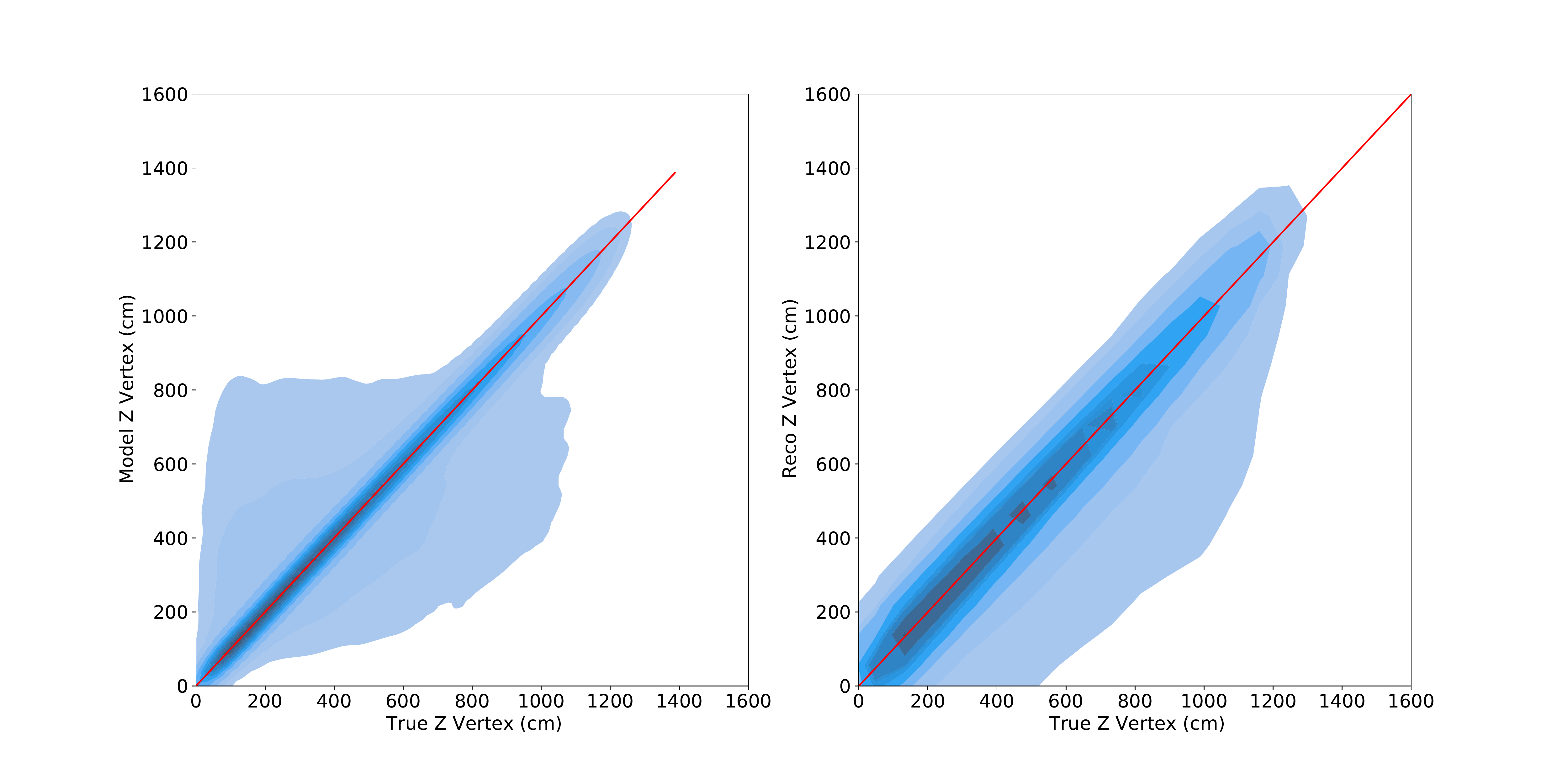}
    \caption[RegCNN model predictions and reconstruction vs true values along z-axis.]{\label{fig:z-full-modelpred-true-vs-reco-kde}RegCNN model predictions and reconstruction vs true values along z-axis.}
  \end{figure}
\addtocontents{lof}{\vspace{\normalbaselineskip}}
\bigskip

\newpage
\begin{figure}[H]
\centering
\begin{subfigure}{0.3\textwidth}
\centering
\includegraphics[width = 1\textwidth]{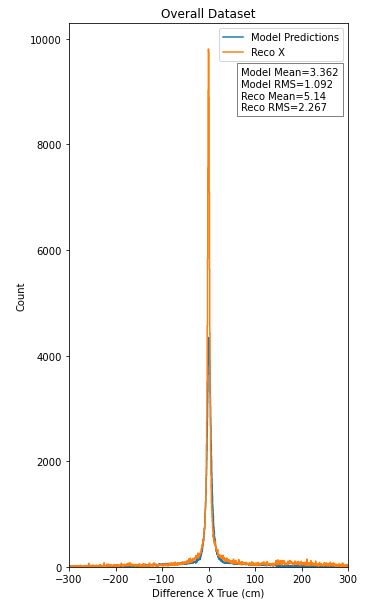}
\caption{\label{fig:x-full-diff-reco-true}}
\end{subfigure}
\begin{subfigure}{0.3\textwidth}
\centering
\includegraphics[width = 1\textwidth]{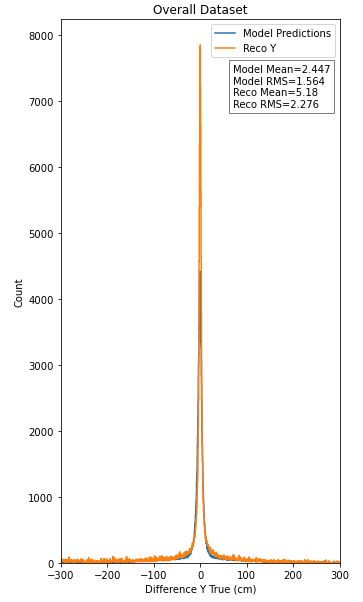}
\caption{\label{fig:y-full-diff-reco-true}}
\end{subfigure}
\begin{subfigure}{0.3\textwidth}
\centering
\includegraphics[width = 1\textwidth]{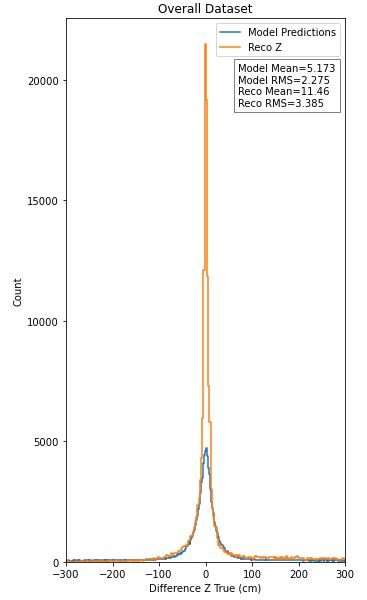}
\caption{\label{fig:z-full-diff-reco-true}}
\end{subfigure}
\caption[Statistical deviation of the two methods from the true vertex values.]{\label{fig:diff_plots} Delta plot of reconstruction and model prediction values from the true values along (a) x-axis (b) y-axis (c) z-axis. Note: Difference in bin values in both reconstruction and model prediction delta plots exist to prevent distortion.}
\end{figure}
\addtocontents{lof}{\vspace{\normalbaselineskip}}

\newpage
\subsection{Comparative Analysis and Conclusion}

In this section, the results are presented in the form of interaction mode and interaction charge. Results are compared with the mean values of the CNN model predictions and that of the traditional reconstruction methods in all axes. The root mean square plots emphasize error in the two methods relative to the true vertex values. The CNN model shows significantly improved statistical results in particular interaction conditions. For instance, vertex predictions for CC interactions will likely out-perform NC interactions. As well, RES interaction mode will likely under-perform compared to QES and DIS interaction modes. The full comparative results are tabulated for the reader in tables \ref{table:qes-results}, \ref{table:dis-results}, \ref{table:res-results} and appendix B.

It is worth reiterating that the model would likely yield better results when trained and validated on similar interaction modes. Another interesting observation to note is that the CNN model's performance would likely suffer when < 5 hits are depicted in one or both views of the CVN pixel map. This is because the CNN model would not be able to detect sufficient features for an accurate prediction. Through appropriate HDF5 data cuts, the model could be applied more effectively in vertexing applications. The CNN model in rock interactions (neutrino interactions with primary vertices outside the near detector boundary) has shown to be more effective in detecting primary vertices. With this knowledge, the CNN model could be applied to help detect interactions outside the boundary of the near detector and assist in the practical application of eliminating these events as a form of an event filter. The results shown below are promising for future work in the application of computer vision related deep learning models in high energy physics.

\newpage
\begin{table}[ht]
\centering 
\caption[QES Results Comparison Table]{QES Results Comparison Table} 
\label{table:qes-results} 
\begin{tabular}{|l||*{4}{c|}}\hline
\backslashbox{Coordinate}{QES Dataset}
&\makebox[6em]{CNN Mean}&\makebox[6em]{CNN RMS}&\makebox[6em]{Reco Mean}
&\makebox[6em]{Reco RMS}\\\hline\hline
X-Axis & 1.04 & 1.02 & 5.27 & 2.295\\\hline
Y-Axis & 2.024 & 1.423 & 5.35 & 2.313\\\hline
Z-Axis & 5.643 & 2.337 & 11.93 & 3.454\\\hline
\end{tabular}
\end{table}
\addtocontents{lot}{\vspace{\normalbaselineskip}}
\bigskip
\begin{table}[ht]
\centering 
\caption[DIS Results Comparison Table]{DIS Results Comparison Table} 
\label{table:dis-results} 
\begin{tabular}{|l||*{4}{c|}}\hline
\backslashbox{Coordinate}{DIS Dataset}
&\makebox[6em]{CNN Mean}&\makebox[6em]{CNN RMS}&\makebox[6em]{Reco Mean}
&\makebox[6em]{Reco RMS}\\\hline\hline
X-Axis & 0.937 & 0.885 & 5.11 & 2.261\\\hline
Y-Axis & 2.113 & 1.454 & 5.16 & 2.271\\\hline
Z-Axis & 3.987 & 1.996 & 11.39 & 3.375\\\hline
\end{tabular}
\end{table}
\addtocontents{lot}{\vspace{\normalbaselineskip}}
\bigskip
\begin{table}[ht]
\centering 
\caption[RES Results Comparison Table]{RES Results Comparison Table} 
\label{table:res-results} 
\begin{tabular}{|l||*{4}{c|}}\hline
\backslashbox{Coordinate}{RES Dataset}
&\makebox[6em]{CNN Mean}&\makebox[6em]{CNN RMS}&\makebox[6em]{Reco Mean}
&\makebox[6em]{Reco RMS}\\\hline\hline
X-Axis & 1.525 & 0.99 & 5.12 & 2.263\\\hline
Y-Axis & 2.948 & 1.717 & 5.14 & 2.267\\\hline
Z-Axis & 5.715 & 2.391 & 11.27 & 3.356\\\hline
\end{tabular}
\end{table}
\addtocontents{lot}{\vspace{\normalbaselineskip}}

\newpage
\begin{figure}[H]
  \centering
    \includegraphics[width=5.5in]{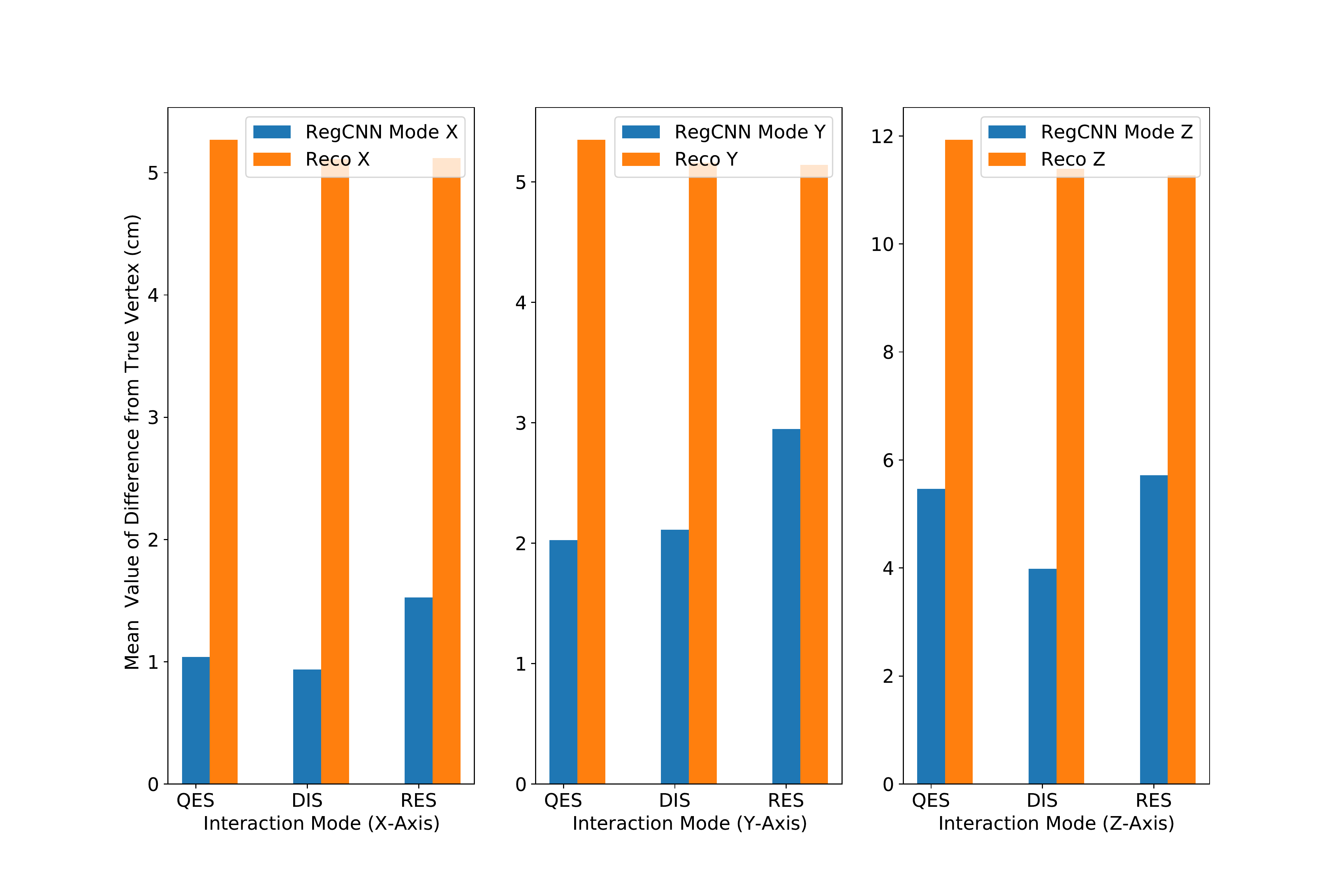}
    \caption[Comparative review of the model performance with respect to interaction mode.]{\label{fig:interaction_mode_compare_mean}Comparative review of the model performance with respect to interaction mode (mean difference value).}
  \end{figure}
\addtocontents{lof}{\vspace{\normalbaselineskip}}
\begin{figure}[H]
  \centering
    \includegraphics[width=5.5in]{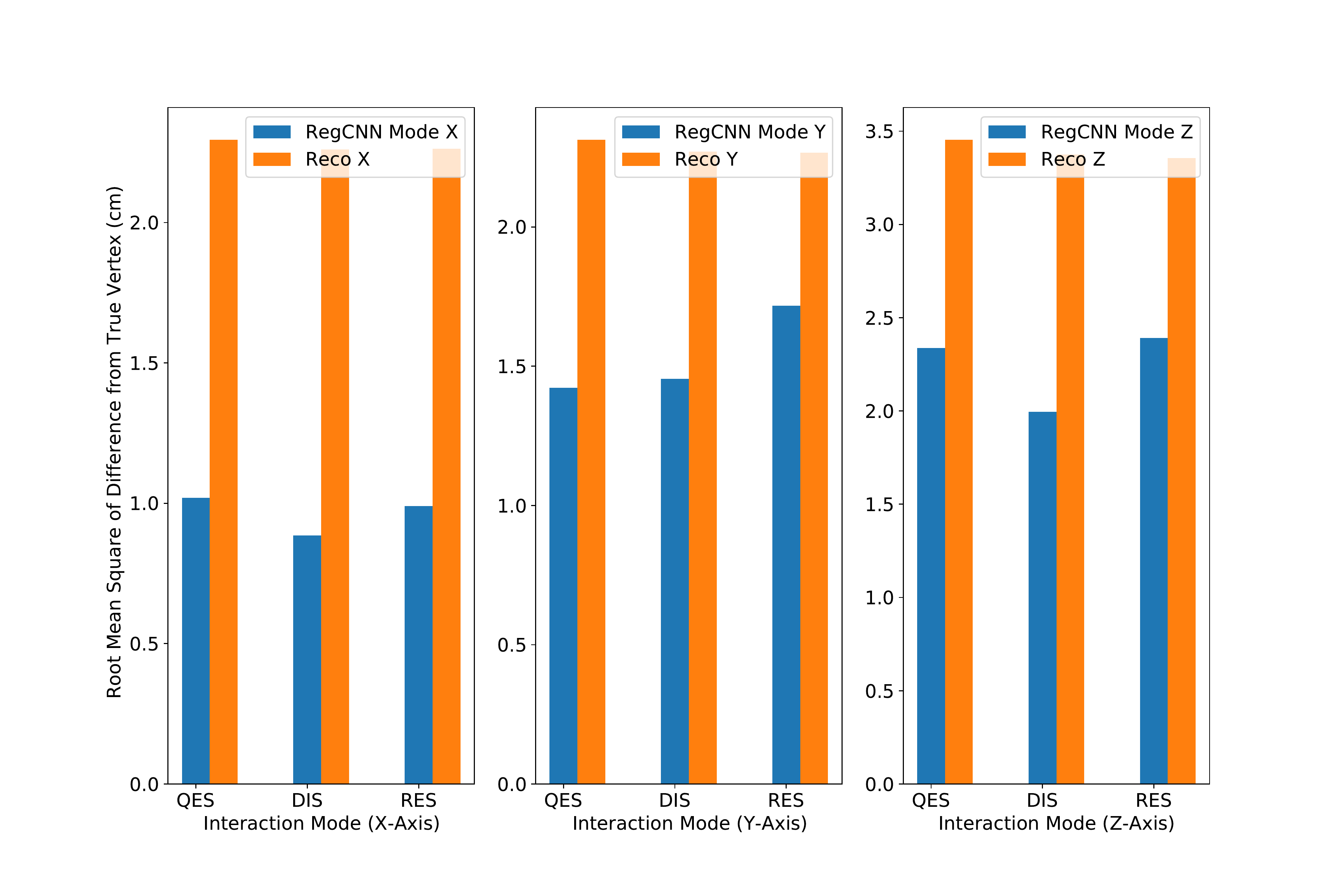}
    \caption[Comparative review of the model performance with respect to interaction mode.]{\label{fig:interaction_mode_compare_rms}Comparative review of the model performance with respect to interaction mode (root mean square of the difference value).}
  \end{figure}
\addtocontents{lof}{\vspace{\normalbaselineskip}}
\begin{figure}[H]
  \centering
    \includegraphics[width=5.5in]{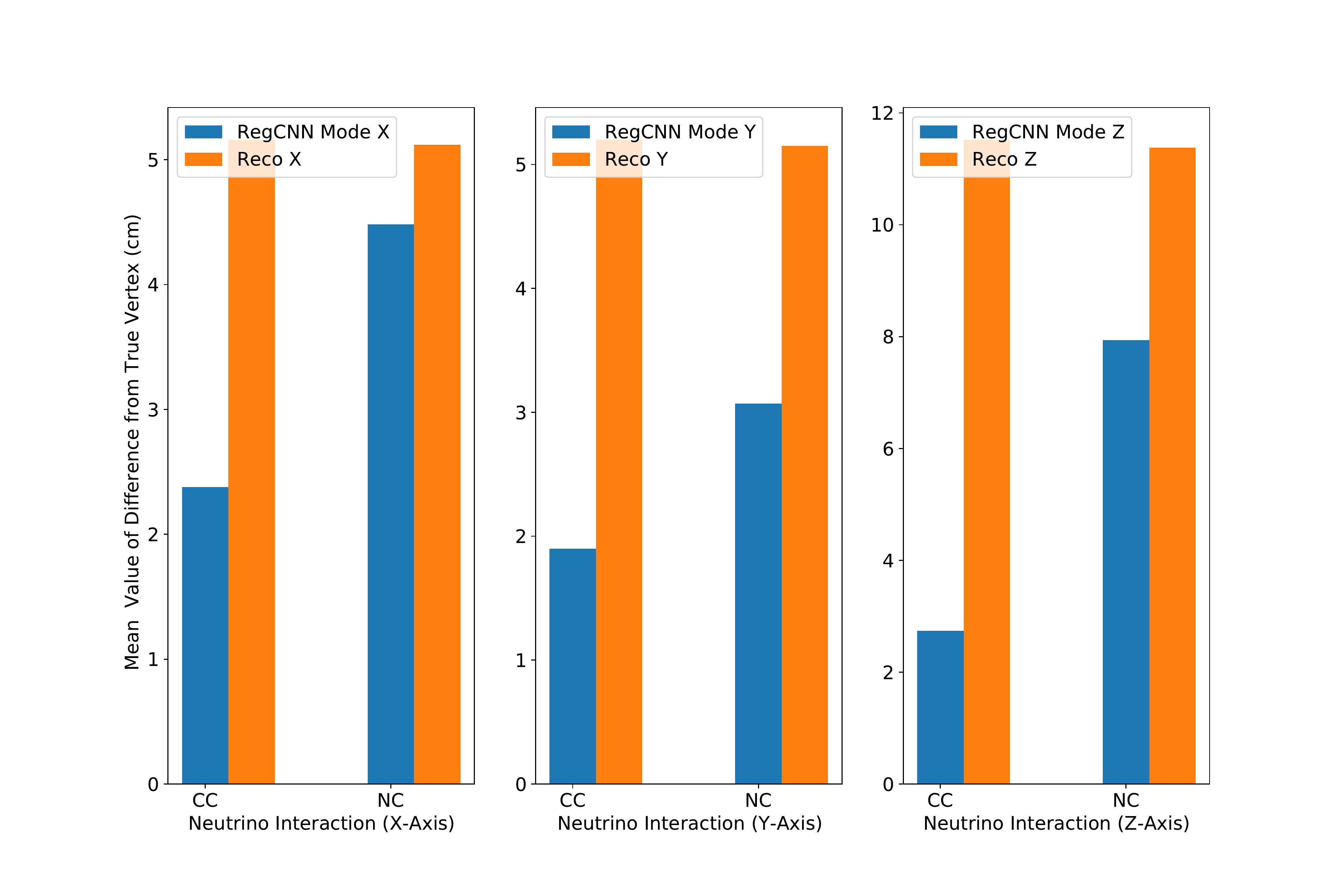}
    \caption[Comparative review of the model performance with respect to neutrino interaction charge (Mean difference value).]{\label{fig:interaction_charge_compare_mean}Comparative review of the model performance with respect to neutrino interaction charge (mean difference value).}
  \end{figure}
\addtocontents{lof}{\vspace{\normalbaselineskip}}
\begin{figure}[H]
  \centering
    \includegraphics[width=5.5in]{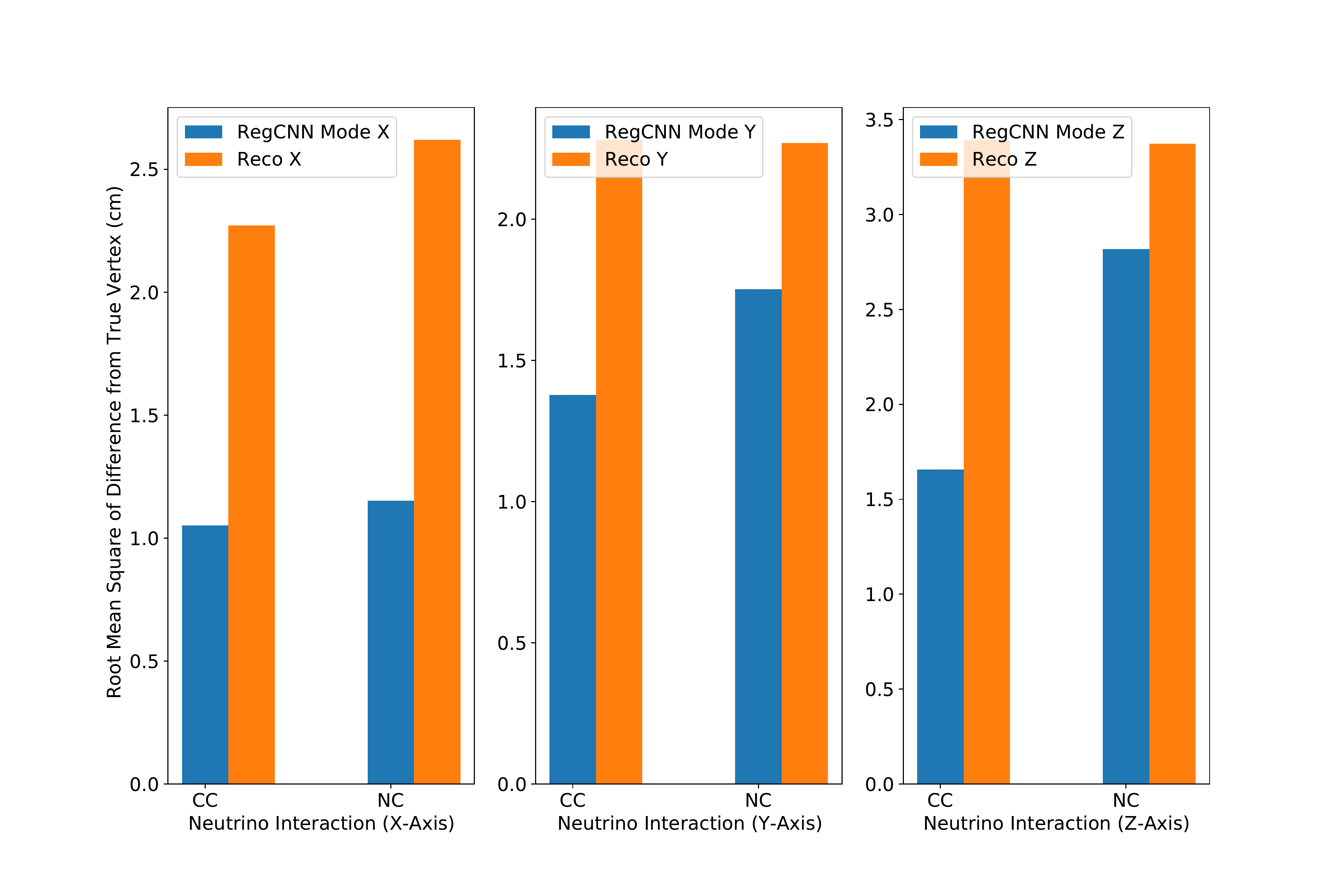}
    \caption[Comparative review of the model performance with respect to neutrino interaction charge (Root mean square of the difference value).]{\label{fig:interaction_charge_compare_rms}Comparative review of the model performance with respect to neutrino interaction charge (root mean square of the difference value).}
  \end{figure}
 \addtocontents{lof}{\vspace{\normalbaselineskip}}

\newpage
\begin{center}
\vspace*{\fill}
\addcontentsline{toc}{section}{BIBLIOGRAPHY}
\section*{\normalfont BIBLIOGRAPHY}
\vspace*{\fill}
\end{center}
\newpage
\let\oldaddcontentsline\addcontentsline
\renewcommand{\addcontentsline}[3]{}
\renewcommand\bibname{\normalfont BIBLIOGRAPHY}
\bibliography{biblio.bib}
\let\addcontentsline\oldaddcontentsline

\newpage
\appendix
\addcontentsline{toc}{section}{APPENDIXES}
\addtocontents{toc}{\vspace{\normalbaselineskip}}
\renewcommand\thesubsection{\Alph{subsection}}
\newpage

\begin{center}
\vspace*{\fill}
\section*{\normalfont APPENDIXES}
\vspace*{\fill}
\end{center}

\newpage
\section*{\normalfont APPENDIX A} \label{App:A}
\section*{Extended Model Prediction Plots} 
\addcontentsline{toc}{subsection}{A. Extended Model Prediction Plots}
\addtocontents{toc}{\vspace{\normalbaselineskip}}

\begin{figure}[H]
  \centering
    \includegraphics[width=3in]{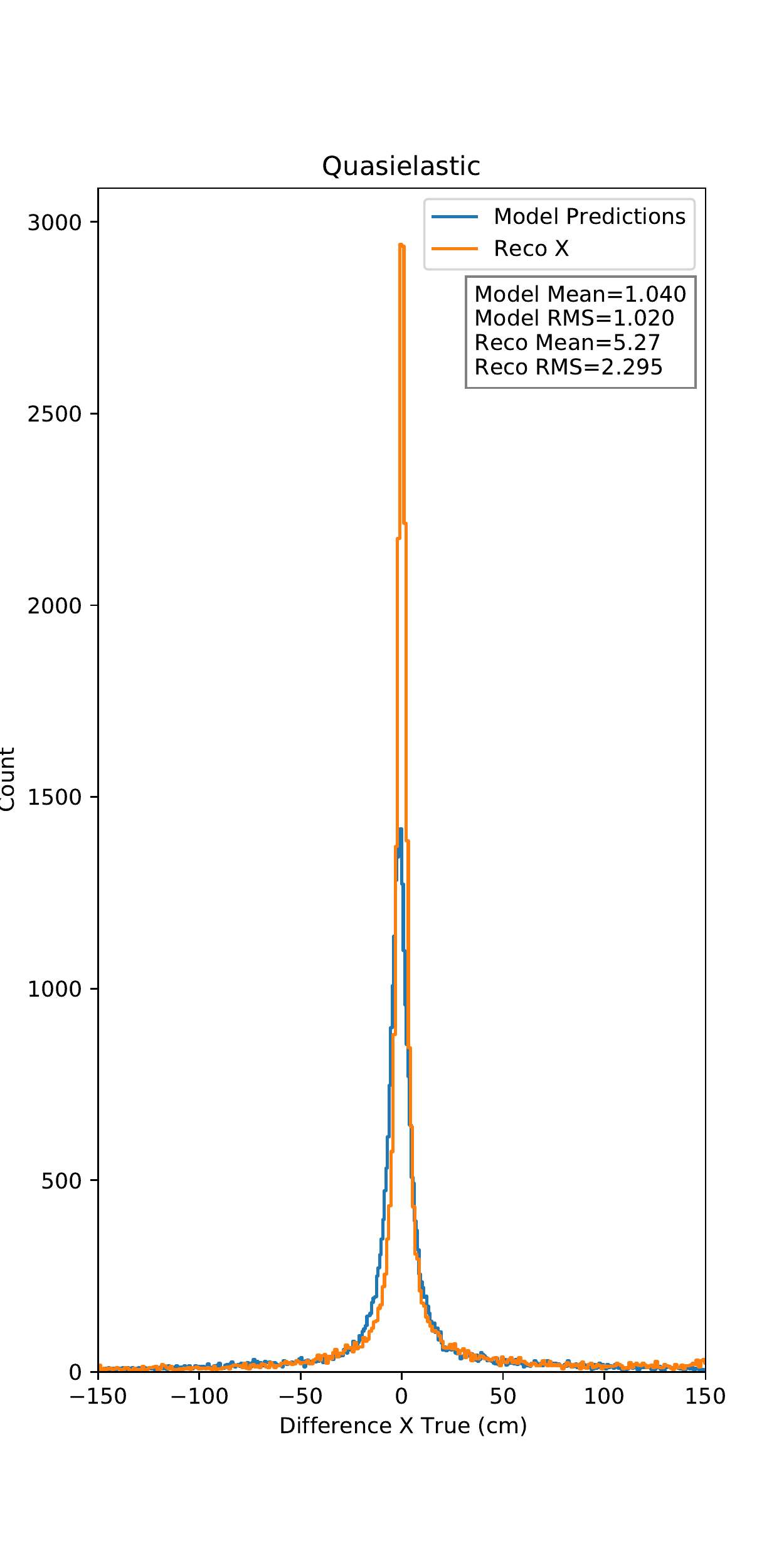}
    \caption[Statistical deviation of the two methods from the true X vertex values.]{\label{fig:x-qes-diff-reco-true}Statistical deviation of the CNN and reconstruction methods from the true X vertex values for QES interaction mode.}
  \end{figure}
\addtocontents{lof}{\vspace{\normalbaselineskip}}

\begin{figure}[H]
  \centering
    \includegraphics[width=3in]{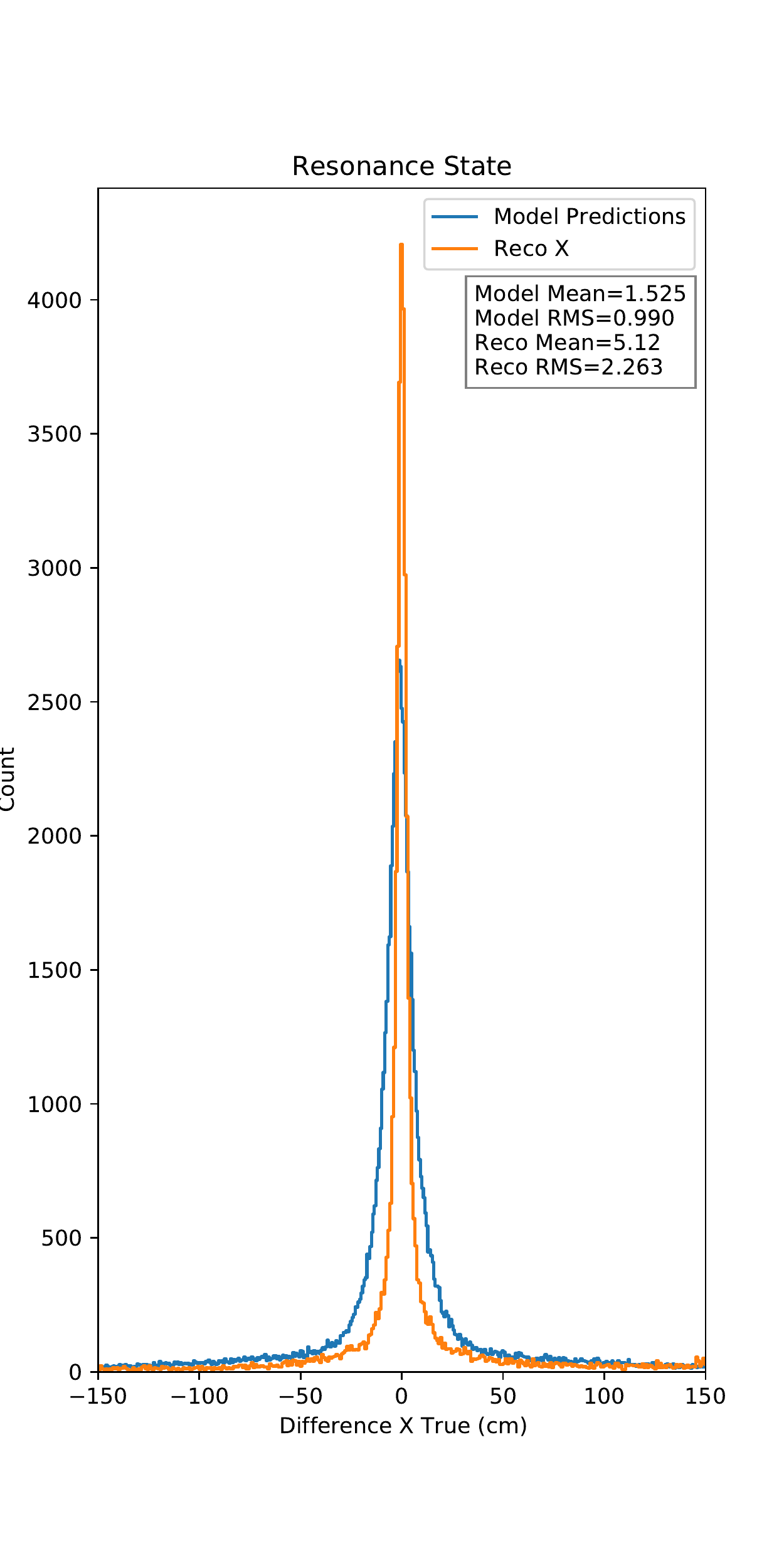}
    \caption[Statistical deviation of the two methods from the true X vertex values..]{\label{fig:x-res-diff-reco-true}Statistical deviation of the CNN and reconstruction methods from the true X vertex values for RES interaction mode}
  \end{figure}
\addtocontents{lof}{\vspace{\normalbaselineskip}}

\begin{figure}[H]
  \centering
    \includegraphics[width=3in]{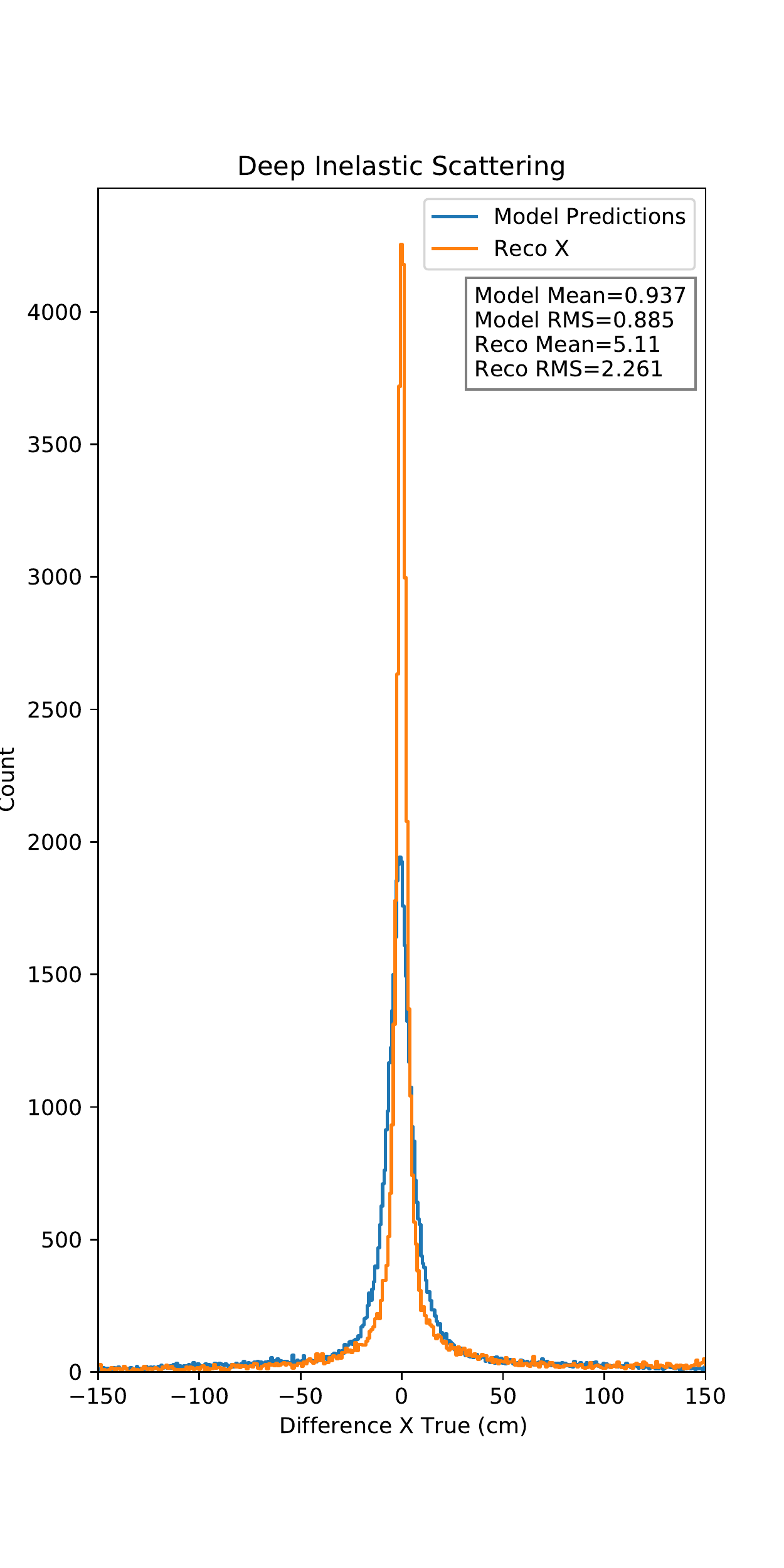}
    \caption[Statistical deviation of the two methods from the true X vertex values.]{\label{fig:x-dis-diff-reco-true}Statistical deviation of the CNN and reconstruction methods from the true X vertex values for DIS interaction mode.}
  \end{figure}
\addtocontents{lof}{\vspace{\normalbaselineskip}}


\begin{figure}[H]
  \centering
    \includegraphics[width=3in]{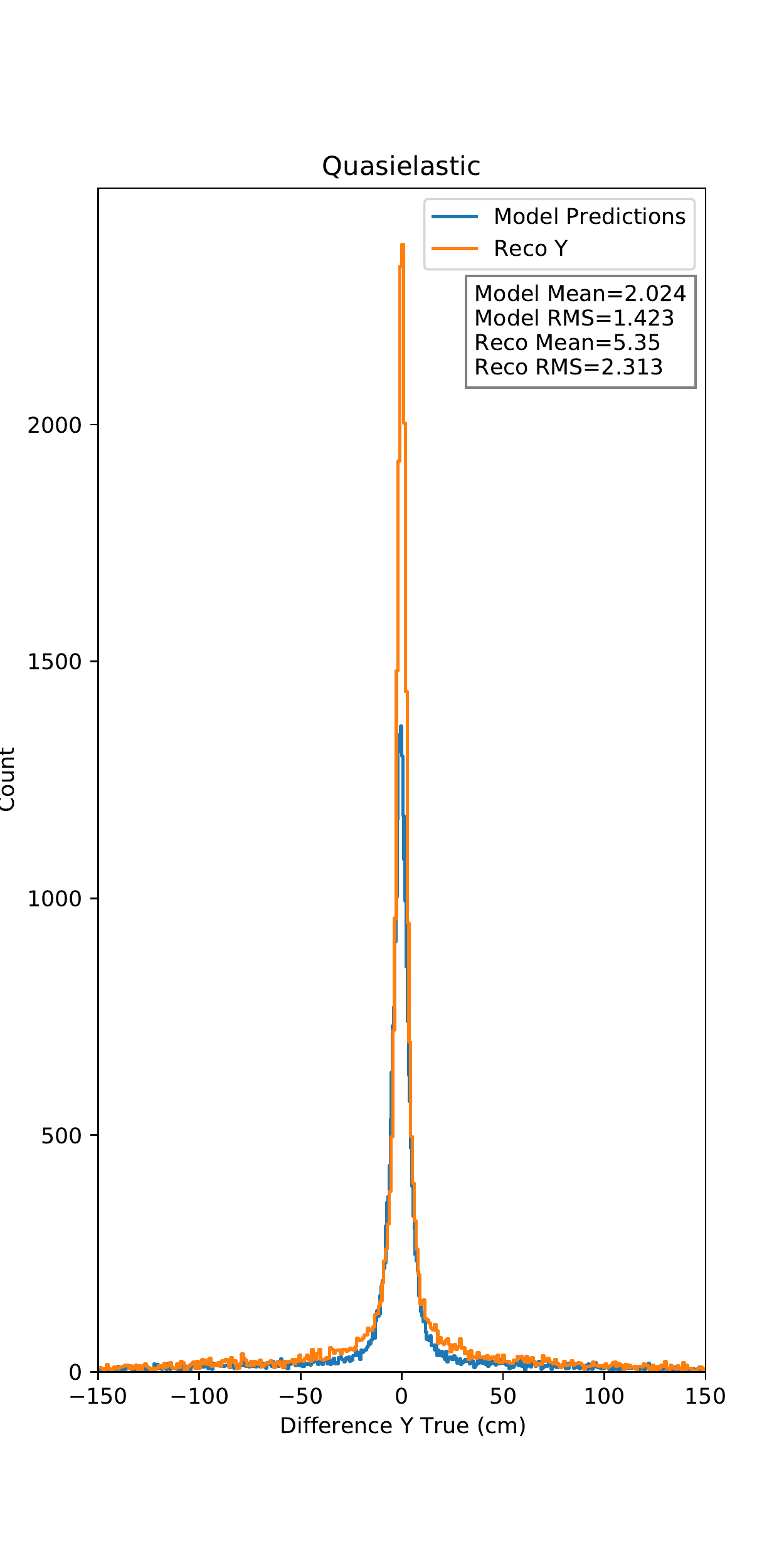}
    \caption[Statistical deviation of the two methods from the true Y vertex values.]{\label{fig:y-qes-diff-reco-true}Statistical deviation of the CNN and reconstruction methods from the true Y vertex values for QES interaction mode.}
  \end{figure}
\addtocontents{lof}{\vspace{\normalbaselineskip}}

\begin{figure}[H]
  \centering
    \includegraphics[width=3in]{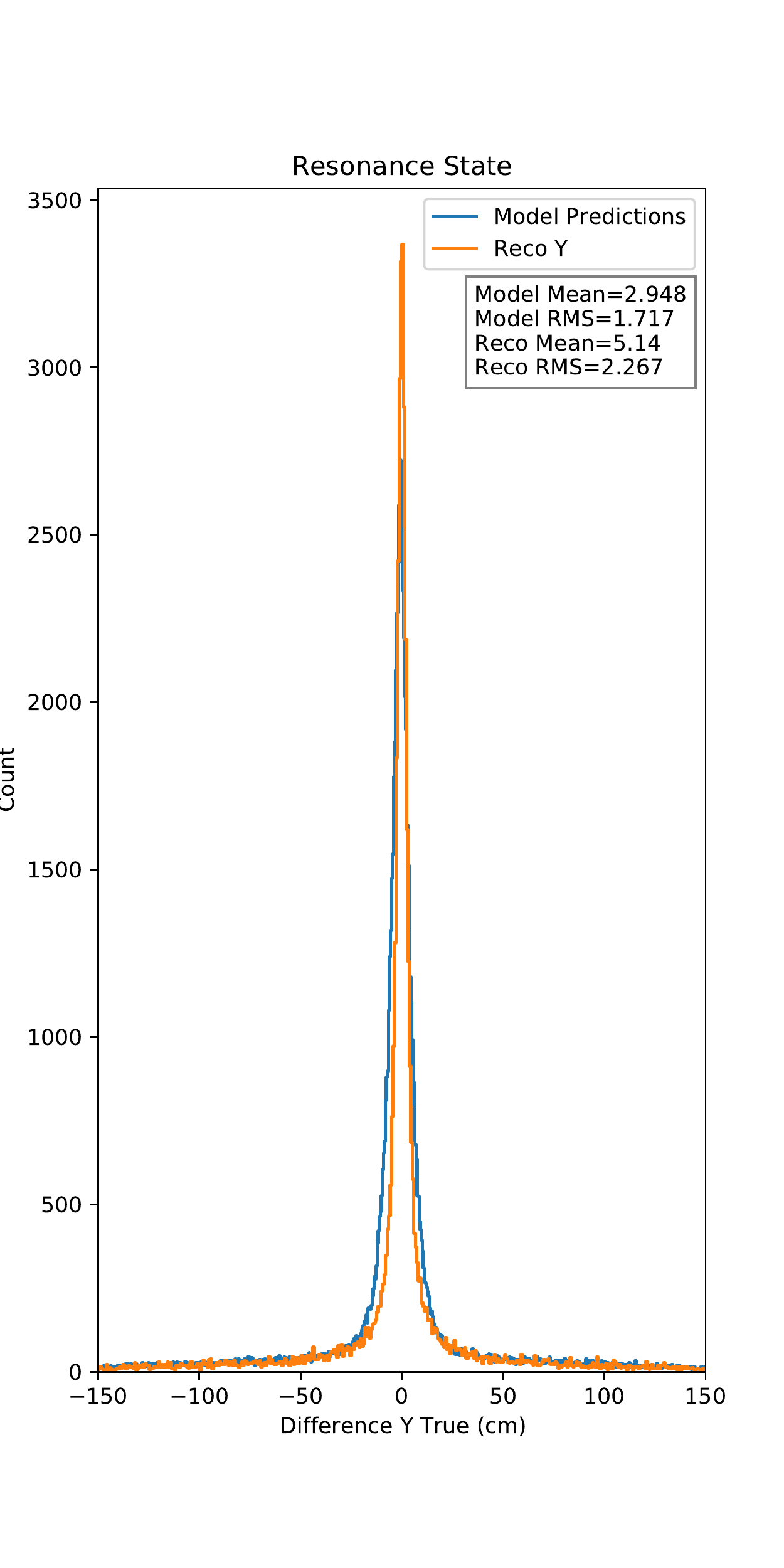}
    \caption[Statistical deviation of the two methods from the true Y vertex values.]{\label{fig:y-res-diff-reco-true}Statistical deviation of the CNN and reconstruction methods from the true Y vertex values for RES interaction mode}
  \end{figure}
\addtocontents{lof}{\vspace{\normalbaselineskip}}

\begin{figure}[H]
  \centering
    \includegraphics[width=3in]{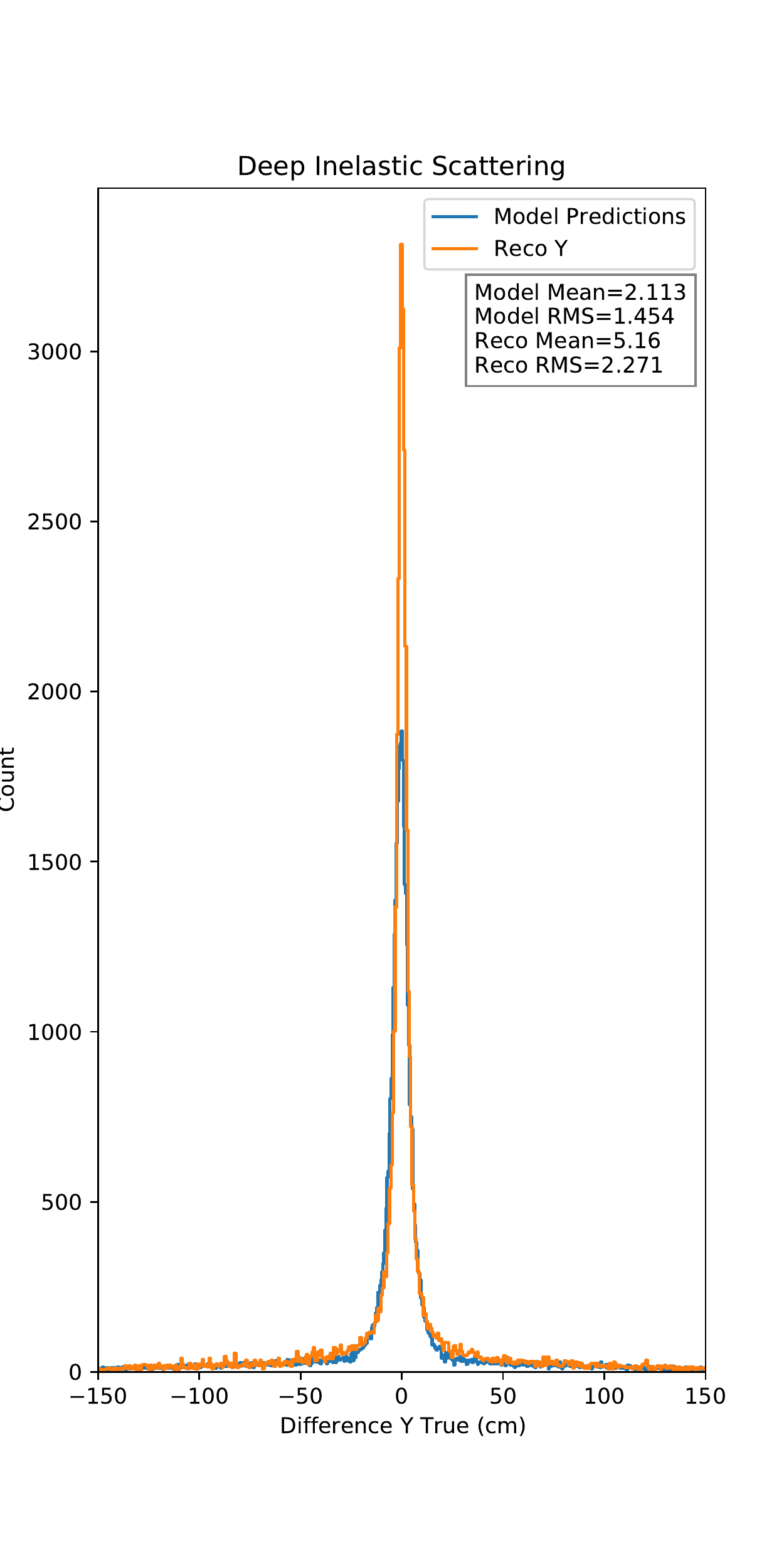}
    \caption[Statistical deviation of the two methods from the true Y vertex values.]{\label{fig:y-dis-diff-reco-true}Statistical deviation of the CNN and reconstruction methods from the true Y vertex values for DIS interaction mode.}
  \end{figure}
\addtocontents{lof}{\vspace{\normalbaselineskip}}


\begin{figure}[H]
  \centering
    \includegraphics[width=3in]{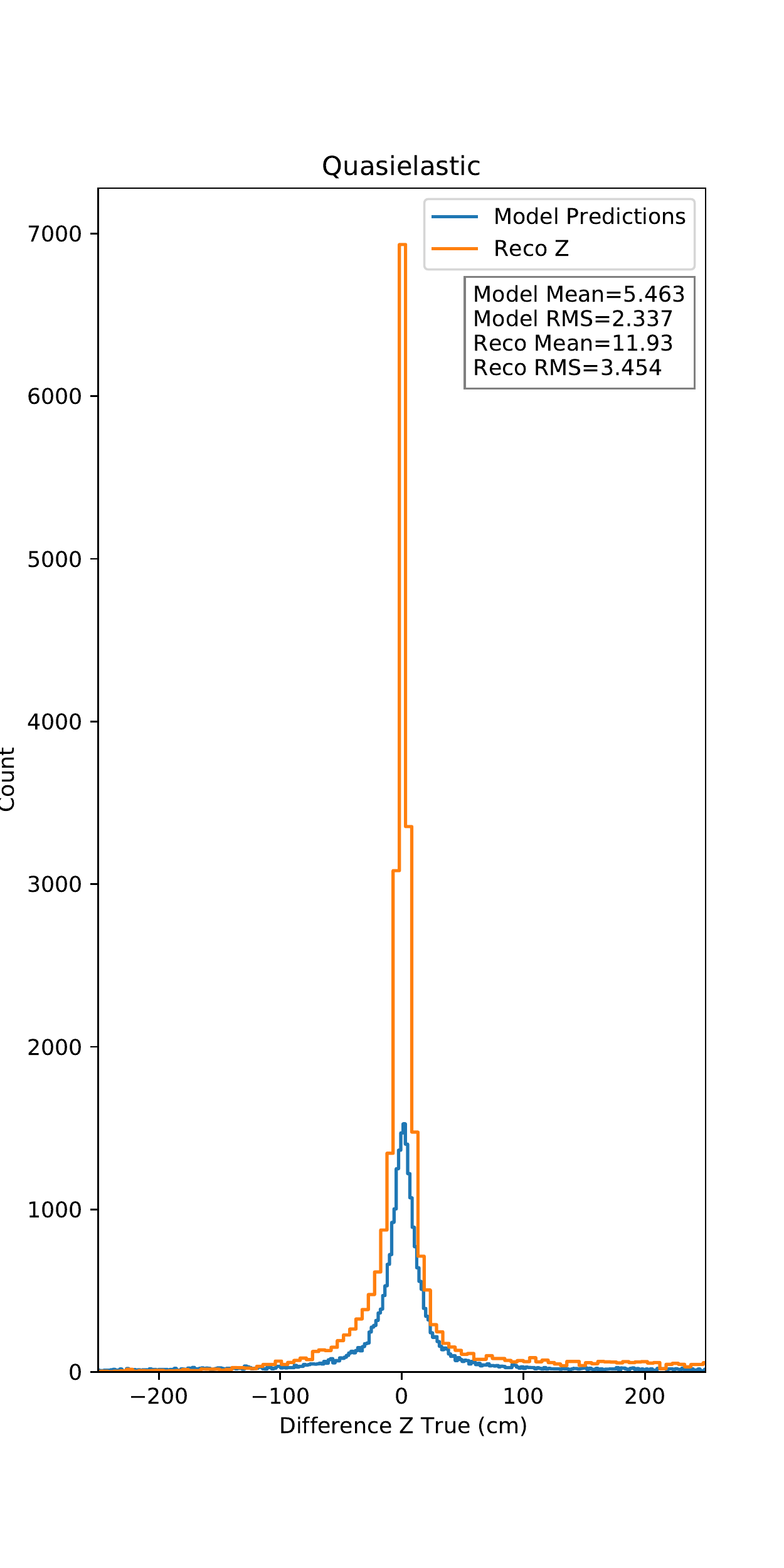}
    \caption[Statistical deviation of the two methods from the true Z vertex values.]{\label{fig:z-qes-diff-reco-true}Statistical deviation of the CNN and reconstruction methods from the true Z vertex values for QES interaction mode.}
  \end{figure}
\addtocontents{lof}{\vspace{\normalbaselineskip}}

\begin{figure}[H]
  \centering
    \includegraphics[width=3in]{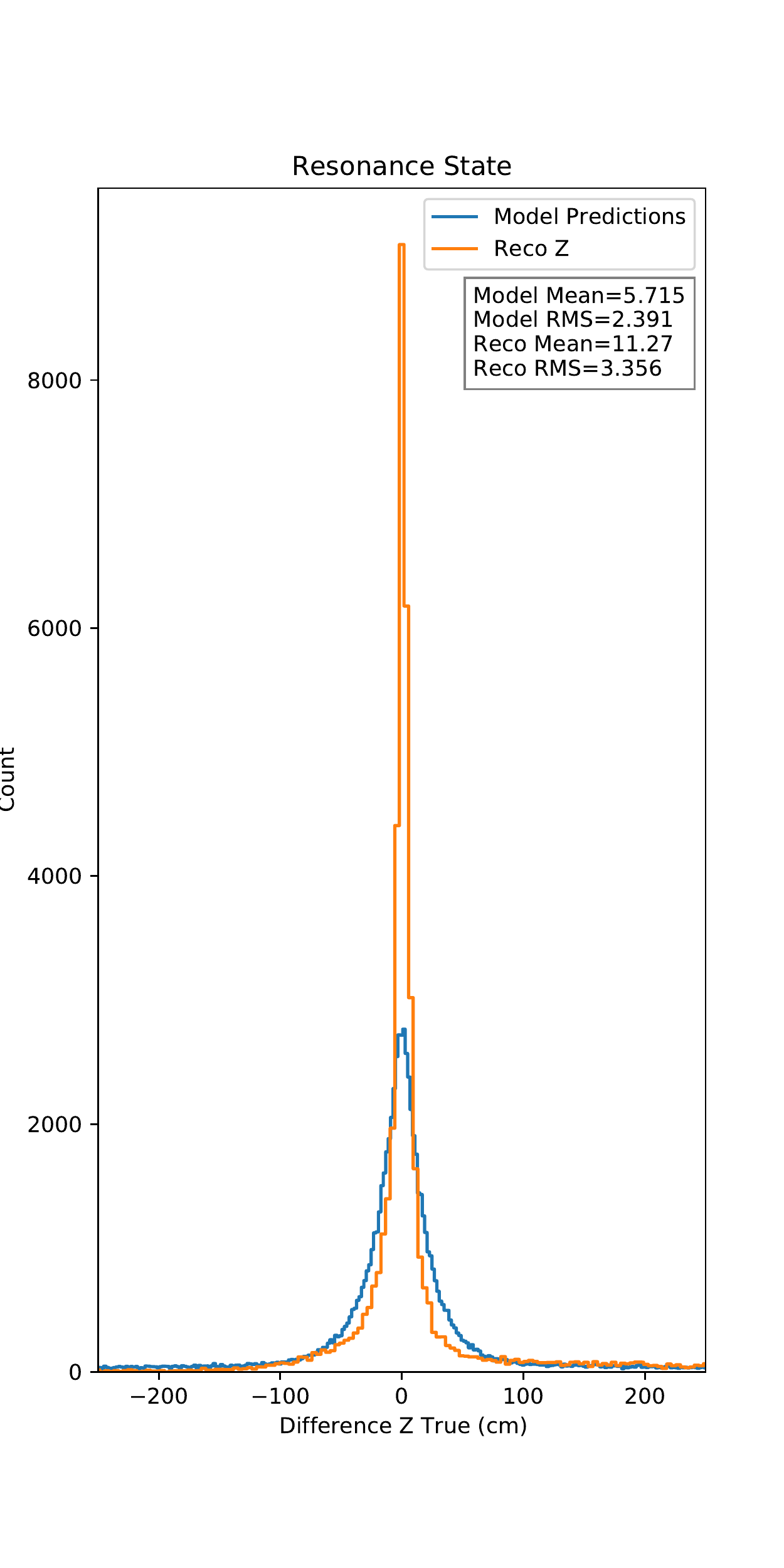}
    \caption[Statistical deviation of the two methods from the true Z vertex values..]{\label{fig:z-res-diff-reco-true}Statistical deviation of the CNN and reconstruction methods from the true Z vertex values for RES interaction mode}
  \end{figure}
\addtocontents{lof}{\vspace{\normalbaselineskip}}

\begin{figure}[H]
  \centering
    \includegraphics[width=3in]{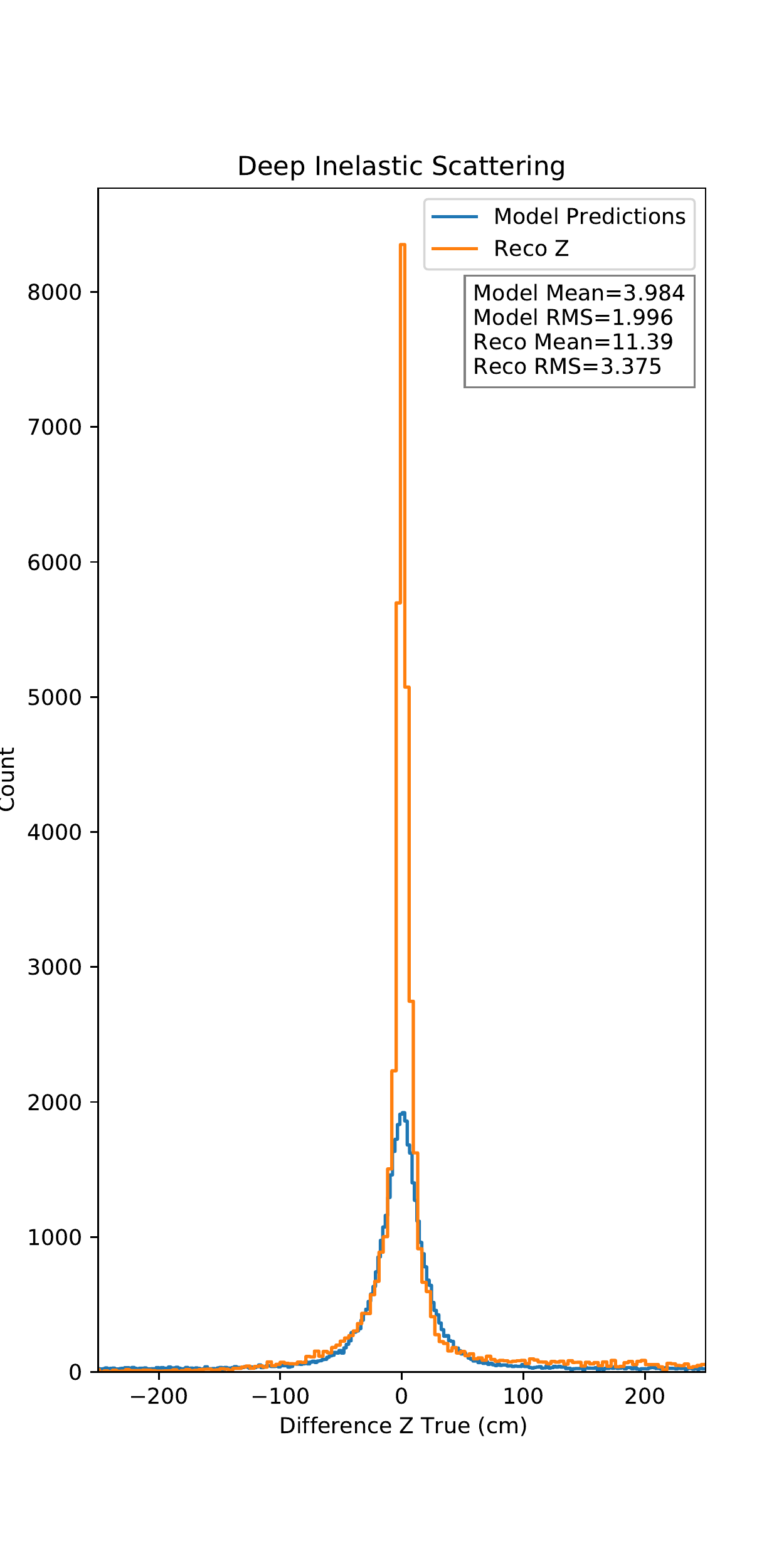}
    \caption[Statistical deviation of the two methods from the true Z vertex values.]{\label{fig:z-dis-diff-reco-true}Statistical deviation of the CNN and reconstruction methods from the true Z vertex values for DIS interaction mode.}
  \end{figure}
\addtocontents{lof}{\vspace{\normalbaselineskip}}

\newpage
\section*{\normalfont APPENDIX B} \label{App:B}
\section*{Additional Comparison Tables} 
\addcontentsline{toc}{subsection}{B. Additional Comparison Tables}
\addtocontents{toc}{\vspace{\normalbaselineskip}}

\begin{table}[ht]
\centering 
\caption[Overall Dataset Results Comparison Table]{Overall Dataset Results Comparison Table} 
\label{table:full-results} 
\begin{tabular}{|l||*{4}{c|}}\hline
\backslashbox{Coordinate}{Full Dataset}
&\makebox[6em]{CNN Mean}&\makebox[6em]{CNN RMS}&\makebox[6em]{Reco Mean}
&\makebox[6em]{Reco RMS}\\\hline\hline
X-Axis & 3.362&1.092&5.14&2.267\\\hline
Y-Axis & 2.447&1.564&5.18&2.276\\\hline
Z-Axis & 5.173&2.275&11.46&3.385\\\hline
\end{tabular}
\end{table}
\addtocontents{lot}{\vspace{\normalbaselineskip}}

\begin{table}[ht]
\centering 
\caption[CC Results Comparison Table]{CC Results Comparison Table} 
\label{table:CC-results} 
\begin{tabular}{|l||*{4}{c|}}\hline
\backslashbox{Coordinate}{CC Dataset}
&\makebox[6em]{CNN Mean}&\makebox[6em]{CNN RMS}&\makebox[6em]{Reco Mean}
&\makebox[6em]{Reco RMS}\\\hline\hline
X-Axis & 2.378&1.052&5.16&2.272\\\hline
Y-Axis & 1.899&1.378&5.2&2.281\\\hline
Z-Axis & 2.741&1.656&11.52&3.394\\\hline
\end{tabular}
\end{table}
\addtocontents{lot}{\vspace{\normalbaselineskip}}

\begin{table}[ht]
\centering 
\caption[NC Results Comparison Table]{NC Results Comparison Table} 
\label{table:NC-results} 
\begin{tabular}{|l||*{4}{c|}}\hline
\backslashbox{Coordinate}{NC Dataset}
&\makebox[6em]{CNN Mean}&\makebox[6em]{CNN RMS}&\makebox[6em]{Reco Mean}
&\makebox[6em]{Reco RMS}\\\hline\hline
X-Axis & 4.482&1.152&5.12&2.262\\\hline
Y-Axis & 3.07&1.752&5.15&2.27\\\hline
Z-Axis & 7.94&2.818&11.38&3.373\\\hline
\end{tabular}
\end{table}
\addtocontents{lot}{\vspace{\normalbaselineskip}}

}

\end{flushleft}
\end{document}